\documentclass{mn2e}
\usepackage{epsfig}
\usepackage{rotating}

\newif\ifAMStwofonts
\AMStwofontstrue

\title{ASTRO-F - SUPER-IRAS - The All Sky Infra-Red Survey}
\author[C.~P.~Pearson et al]
       {Chris P. Pearson$^1$$^{,5}$\thanks{Further information contact Chris Pearson (cpp@ic.ac.uk)         $http://astro.ic.ac.uk/\sim cpp/astrof/$ },
H.~Shibai$^2$, 
T.~Matsumoto$^3$, 
H.~Murakami$^3$,
T.~Nakagawa$^3$,  
\vspace*{0.3cm}
\\ {\LARGE \textup{ 
M.~Kawada$^2$,
T.~Onaka$^4$, 
H.~Matsuhara$^3$, 
T.~Kii$^3$,
I.~Yamamura$^3$, 
T.~Takagi$^1$ 
}
\vspace*{0.1cm}
}\\
        $^1$ Astrophysics Group, Imperial College of Science Technology and Medicine, Blackett Laboratory, Prince Consort Rd., London,SW72BW,U.K. \\
        $^2$ Infrared Astrophysics laboratory, Nagoya University, Furo-cho, Chikusa-ku, Nagoya 464-8602, Japan\\
       $^3$ Institute of Space and Astronautical Science, Yoshinodai 3-1-1, Sagamihara, Kanagawa 229 8510, Japan\\
        $^4$ Department of Astronomy, School of Science, University of Tokyo, Tokyo 113-0033, Japan\\    
        $^5$ Centre for Astrophysics and Planetary Science, University of Kent at Canterbury, Canterbury, Kent, U.K.}     

\date{Accepted .\\
      Received ;\\
      in original form 2002 October }
\pagerange{\pageref{firstpage}--\pageref{lastpage}}

\begin{document}

\label{firstpage}

\maketitle

\begin{abstract}
We review the next generation Japanese infrared space mission, ASTRO-F. ASTRO-F will be the first survey of the entire sky at infrared wavelengths since the IRAS mission almost 20 years ago. ASTRO-F will survey the entire sky in 4 far-infrared bands from 50-200microns and 2 mid-infrared bands at 9 and 20microns to sensitivities of 10-1000 times deeper than the IRAS satellite at angular resolutions of 25-45arcsec (c.f. IRAS 2-5arcmins). ASTRO-F can be considered a SUPER-IRAS. Using the galaxy evolution model of Pearson (2001) we produce expected numbers of sources under 3 different cosmological world models. We predict that ASTRO-F will detect of the order of 10's millions of sources in the far-infrared wavelength bands, most of which will be dusty LIG/ULIGs of which as many as half will lie at redshifts greater than unity. We produce number-redshift distributions, flux-redshift and colour-colour diagrams for the survey and discuss various segregation and photometric redshift techniques. Furthermore, we investigate the large scale structure scales that will be accessed by ASTRO-F, discovering that ASTRO-F and SIRTF-SWIRE probe both different scales and redshift domains and concluding that the 2 missions will supplement rather than supplant one another.

\end{abstract}

\bigskip

\begin{keywords}
Cosmology: source counts -- Infrared: source counts, Surveys -- Galaxies: evolution.
\end{keywords}

\bigskip

\section{Introduction}\label{sec:introduction}
	In order to understand such fundamental cosmological questions such as type dependent galaxy evolution, star formation history, and the large scale structure of Universe, large data sets of homogeneous data are required. Most recently, these large data sets have been mainly accumulated at NIR/optical wavelengths (e.g. The Sloan Digital Sky Survey, SDSS, $\approx 10^{6}$ galaxies over 1/4 of the sky, ~\cite{stoughton02}, The 2 degree field galaxy redshift survey, 2dFGRS, $\approx 250,000$ galaxies over  $\approx$2000sq.deg. of the sky ~\cite{colless01}). To investigate galaxy evolution and starformation in the high redshift z$>$1 Universe, due to finite observation times, a trade off must be made between the area and the depth of a given survey, that is to say, the greater the sensitivity or depth required, the narrower the survey has to be, e.g. The Hubble and SUBARU Deep Fields, Williams et al. ~\shortcite{williams96}, Maihara et al. ~\shortcite{maihara01}. Hence, although the two HDF fields have produced excellent images and some $\sim$1000 spectroscopic redshifts with $>$30 galaxies at z$>$2 (Cohen et al. ~\shortcite{cohen00}, Dennefeld  ~\shortcite{dennefeld00}, they are effectively limited to 2 lines of sight (HDF-N, 5.3sq.arcmin., Williams et al. ~\shortcite{williams96} and HDF-S, 0.7sq.arcmin.,  Williams et al. ~\shortcite{williams00}, Gardner et al. ~\shortcite{gardner00}) and are therefore somewhat susceptible to statistical uncertainty (e.g. from large scale structure). For example, almost one quarter of the total z$<$1.1 0.65$\umu$m luminosity comes from just 4 galaxies.

In the past few years considerable progress has been made in the understanding of the star formation rate of high redshift galaxies using the Lyman Drop Out technique to isolate and identify galaxies at redshifts of $\approx$2$\sim$4 (Steidel et al. ~\shortcite {steidel96}, Madau et al. \shortcite{madau96}). Well over 500 Lyman-break systems have now been observed at z$>$3. These are UV-bright, blue spectra, actively star-forming galaxies. Estimations of star formation rates of a few to $\sim 50 M_{\sun}yr^{-1}$ led to the popular Madau Plot ~\cite{madau96} and the somewhat premature implication that the star formation rate of the Universe peaked at z$\approx$1$\sim$2. However, the early analysis of the optical/UV star formation rate neglected the effects due to dust. Subsequent analysis correcting for extinction did indeed imply corrections to the star formation rates of Lyman-break galaxies of factors of 2-7 (Pettini et al. ~\shortcite{pettini98}, Dickinson ~\shortcite{dickenson98}, Meurer et al.  ~\shortcite{meurer99}, Calzetti \& Heckman~\shortcite{calzetti99}, Steidel et al. ~\shortcite{steidel99}, Goldader et al. ~\shortcite{goldader02}).

Indeed, studies with the Infrared Space Observatory, ISO ~\cite{kessler96}, of the Hubble Deep Fields have revealed star formation rates at least comparable to or higher than those of the optical/UV studies (Rowan-Robinson et al.~\shortcite{mrr97}, Mann et al. ~\shortcite{mann02}). At sub-millimetre wavelengths, surveys  with SCUBA on the JCMT ~\cite{holland99} revealed a large ($>$3000/sq.deg. at $S_{850} > 2mJy$) strongly evolving population of sources with bolometric luminosities $>10^{12}L_{\sun}$ and star formation rates of $\sim 300- >1000 M_{\sun}yr^{-1}$ lying at redshifts $>$1 (  Smail et al.~\shortcite{smail97}, Hughes et al.~\shortcite{hugh98}, Barger et al.~\shortcite{barg98}, Blain et al.~\shortcite{blain99}, Barger et al.~\shortcite{barg99}, Dunne et al.~\shortcite{dunne00} Fox et al.~\shortcite{fox02}). The straight forward interpretation of these results is that most massive star formation occurs in dense molecular clouds that obscure the view at optical wavelengths. What is seen at optical/UV wavelengths represents star formation at the periphery of the clouds and is but the tip of the iceberg.

Of course this revelation is not new. The first significant evidence came from the IRAS satellite, launched in 1983, which surveyed almost the entire sky in four wavebands, 100$\umu$m, 60$\umu$m, 25$\umu$m \& 12$\umu$m, detecting more than 25,000 galaxies and providing a huge legacy of data. IRAS revealed a new population of galaxies in the IR indicating that most of the star formation history had in fact been missed by the optical observations. The local ratio of $L_{IR}/ L_{opt}$ was found to vary from $\sim 30\%$ for normal quiescent galaxies (Rowan-Robinson et al.~\shortcite{mrr87}, Corbelli, Salpeter \& Dicky ~\shortcite{corbelli91} to as much as $\sim 90\%$ for luminous $L_{IR} \geq 10^{11}L_{\sun}$ and ultraluminous $L_{IR} \geq 10^{12}L_{\sun}$ galaxies (e.g. Soifer et al. ~\shortcite{soifer86}. 

Further evidence of a darker side to star formation has come from  measurements, detections and upper limits of the integrated extragalactic background light at 2.2$\umu$m, 3.5$\umu$m, 140$\umu$m and 240$\umu$m to 2000$\umu$m by the COBE, ISO and IRTS missions (Puget et al. ~\shortcite{puget96}, Fixsen et al. \shortcite{fixsen98}, Hauser et al. ~\shortcite{hauser98}~\shortcite{hauser01}, Lagache et al. ~\shortcite{lagache00a}, Lagache \& Puget~\shortcite{lagache00b}, Gorijian, Wright \& Chary ~\shortcite{Gorijian00}, Matsumoto et al. ~\shortcite{mats00}, Wright \& Reese ~\shortcite{wright00}, Finkbeiner et al. ~\shortcite{fink00}). These data have indicated that the far-infrared background is comparable to the optical near-infrared (NIR) background, i.e. that 50$\%$ of the integrated rest frame optical/UV emission from hot, young stars is reprocessed by dust and re-radiated at mid-infrared (MIR) and far-infrared (FIR) wavelengths. Therefore the rest frame optical/UV luminosities of galaxies only contribute a lower limit to the star formation history of the Universe ~\cite{chary01}. 

In order to tightly constrain general mass evolution of galaxies and the star formation history of the Universe, large area, wide field infrared surveys to redshifts 1$\sim$2 are required. For example, although the optical/UV studies of Lyman-break galaxies sample large volumes at {\it high} redshift (the volume encompassed at 2$<$z$<$10 is 20x the volume at z$<$1 ,$\Omega =0.3,\Lambda=0.7$) most cosmological evolution takes place at lower redshifts. The ISO and SIRTF ~\cite{rieke00} missions are observatories making surveys of the order of 10's sq.deg.. The IRAS mission produced a pioneering survey of the infrared sky but could only reach out to relatively low redshifts (e.g. z$_{med}\sim$0.03 for the PSCz,~\cite{saun00}) and is now more than 20 years old. To extend unbiased measurements of the star formation rate out to higher redshifts a new all sky infrared survey is required.

ASTRO-F, also known as the Infra-Red Imaging Surveyor (IRIS, \cite{mura98}) will be the second infrared astronomy mission of the Japanese Institute of Space and Astronautical Science (ISAS). ASTRO-F is a 67cm cooled telescope and will be dedicated for all sky and large area surveys in the infrared, one of its' primary objectives being to map the entire sky in 4 FIR bands ranging from 50-200$\umu$m. With its' timely launch almost on the 20th anniversary of the IRAS mission ASTRO-F can indeed be regarded as a {\it SUPER-IRAS}.

In this paper, we review the ASTRO-F mission its objectives and predicted results (see Takeuchi et al. ~\shortcite{take99} for an earlier discussion of the ASTRO-F mission). Note that although ASTRO-F will also provide equally valuable information in the areas of galactic star-formation, planetary systems and innumerable other fields of galactic astronomy and astrophysics, here we will concentrate solely on and its extragalactic all sky survey, as this is one of the primary motivations behind the mission.
In section~\ref{sec:ASTRO-F} we introduce the ASTRO-F mission and the infrared all sky survey. Section~\ref{sec:model} explains the model parameters used to predict the survey predictions. In section~\ref{sec:survey} we present the predictions of the all sky survey and in section~\ref{sec:colours} we discuss the simple methods of colour discrimination of sources in the all sky survey. A summary of results and the conclusions are given in section~\ref{sec:conclusions}.

\bigskip

\begin{figure}
\centering
\centerline{
\psfig{figure=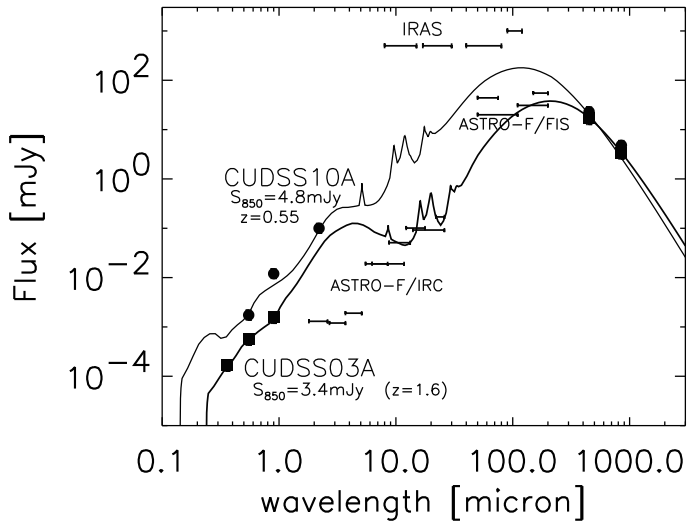,height=8cm}
}
\caption{ASTRO-F survey sensitivities ($5\sigma$) for IRC and FIS instruments assuming All Sky Survey sensitivities for the FIS (FIR wavelengths) and pointing sensitivities from Pearson et al. (2001) for the IRC (MIR wavelengths). IRAS flux limits are also shown for comparison. Sensitivity limits are compared with 2 sources from the Canada-U.K. Deep SCUBA Survey (CUDSS, Lilly et al.(1999)). CUDSS10A has a spectroscopic redshift of 0.55 and a flux at 850$\umu$m of S$_{850}$=4.8mJy. For CUDSS03A (S$_{850}$=3.4mJy), the adopted redshift is 1.6, which gives minimum chi square fit to the SED (Takagi et al. (2002)). Both of these sources would be detected by the IRC. CUDSS10A would also be detected by the FIS in all 4 bands and even CUDSS03A would be detected marginally in the 2 most sensitive FIS bands. 
\label{sensitivity}}
\end{figure}

\section{The ASTRO-F All Sky Survey}\label{sec:ASTRO-F}

	The ASTRO-F infrared space telescope~\cite{mura98} incorporates a 67cm diameter silicon carbide Ritchey-Chretien design telescope (11kg primary mirror) cooled to 5.8K (the detectors to 1.8$\sim$2.5K) by $\approx$170l of liquid Helium contained within a light weight cryostat ~\cite{kaneda02}. The temperature of the outer wall of the cryostat is expected to be maintained below 200K via radiation cooling and a pair of 2-stage Stirling-cycle coolers ensure minimum heat flow from the outer wall of the cryostat, thus almost doubling the lifetime of the Helium providing a mission lifetime of more than 550 days. Furthermore, the use of mechanical coolers means that the near-infrared detectors will still be usable even after Helium exhaustion. ASTRO-F is scheduled to be launched in mid 2004 into a sun-synchronous polar orbit with an altitude of 750 km with ISAS's M-V launch vehicle.

\begin{table*}
\caption{Far Infra-Red Surveyor All Sky Survey Parameters}
\renewcommand{\arraystretch}{1.4}
\setlength\tabcolsep{15pt}
\begin{tabular}{@{}lllll}
\hline\noalign{\smallskip}
Filter & Wavelength Band & Array Size & Pixel size (Diffraction Beam) & $5\sigma$ Sensitivity \\
\noalign{\smallskip}
\hline
\noalign{\smallskip}
N60 & $50-75\mu$m & $20$x$2$ & $26.79''(21.6'')$ & $44$mJy \\
N170 & $150-200\mu$m & $15$x$2$ & $44.20''(60'')$ & $55$mJy  \\
Wide-S & $50-110\umu$m & $20$x$3$ & $26.79''(30'')$ & $20$mJy  \\
Wide-L & $110-200\mu$m & $15$x$3$ & $44.20''(50'')$ & $31$mJy \\
\noalign{\smallskip}
\hline
\noalign{\smallskip}
\end{tabular}\\
\label{FIS}
\end{table*}

\begin{table*}
\caption{Infra-Red Camera All Sky Survey Parameters}
\renewcommand{\arraystretch}{1.4}
\setlength\tabcolsep{15pt}
\begin{tabular}{@{}lllll}
\hline\noalign{\smallskip}
Filter & Wavelength Band & FOV & Spatial (in pixel) Resolution & $5\sigma$ Sensitivity  \\
\noalign{\smallskip}
\hline
\noalign{\smallskip}
IRC-MIR-S9W & $6-11\mu$m & $10'$x$10'$ & $4.7''$x$9.4''$ & $50$mJy\\
IRC-MIR-L20W & $14-26\mu$m & $10'$x$10'$ & $4.7''$x$9.4''$ & $100$mJy\\
\noalign{\smallskip}
\hline
\noalign{\smallskip}
\end{tabular}\\
\label{IRC}
\end{table*}

ASTRO-F incorporates 2 focal plane instruments covering the entire IR region from the K-band to 200$\umu$m. The first focal plane instrument is the Far-Infrared Surveyor (see Table~\ref{FIS} \cite{kaw98} \cite{takahashi00}) which will survey the entire sky simultaneously in 4 FIR bands from 50 to 200$\umu$m with angular resolutions of 25 - 45 arcsec. The FIS has 2 kinds of detector arrays; a short wavelength array (50-110$\umu$m) comprised of unstressed Ge:Ga detectors and a long wavelength array (110-200$\umu$m)comprised of stressed Ge:Ga detectors. These 2 arrays are divided into 2 further narrow and wide bands known as N60, WIDE-S, N170, WIDE-L respectively (see Table~\ref{FIS}). The two arrays are tilted against the scan direction by 26.5deg. such that the interval between the scan paths traced by the detector pixels is half of the physical pitch of the pixels. The long and short wavelength arrays observe almost the same area of sky, the only difference being due to the difference in the size of the arrays. As well as the all sky survey mode the FIS may also be used for pointing observations and includes a Fourier-transform spectrometer mode that can be operated via the wideband arrays with a resolution of $\approx 0.2cm^{-1} (\lambda / \Delta \lambda =250-1000, \lambda =50-200 \umu m$ allowing imaging spectroscopy of selected sources. 

In survey mode, ASTRO-F will perform a continuous scan of the sky. ASTRO-F spins around the Sun pointed axis once every orbit of 100 minutes, keeping the telescope away from the Earth. The result is to trace out a great circle with a solar elongation of 90deg scanning the sky at a rate of 3.6arcmin/sec.

The second focal-plane instrument is the Infrared Camera (IRC, \cite{mat98},\cite{wat00}, \cite{wada02}). The IRC employs large-format detector arrays and will be used to take deep images of selected sky regions in the near-infrared and mid-infrared range after the all sky survey has been performed. The IRC has 3 cameras each with 3 filters. The IRC-NIR has bands in K, L, M plus a grism and prism from 2.5-5$\umu$m \& 2.0-5$\umu$m respectively. The IRC-MIR-S has narrow bands at 7$\umu$m \& 11$\umu$m and a wide band at 9$\umu$m, plus 2 grisms at 5-8$\umu$m \& 7-12$\umu$m. The IRC-MIR-L has narrow bands at 15$\umu$m \& 24$\umu$m, a wide band at 20$\umu$m, plus 2 grisms covering the range 11-19$\umu$m \& 18-26$\umu$m. Pearson et al.~\shortcite{cpp01a} made detailed predictions for small, deep and large, shallow area survey strategies using the IRC instrument in pointing mode. The possibility of adding the 2 wide band channels of the IRC-MIR (S9W, S20W)in scan mode is currently under investigation ~\cite{ishihara02}. In All Sky Survey mode the IRC would utilize just one line (perpendicular to the scan direction of the telescope beam) of the mid-IR detector arrays of the MIR-S and MIR-L channels. Data from these channels would be sampled and downloaded to the ground simultaneously with the FIS all-sky survey data. Typical angular resolutions of $\sim$10\arcsec are expected (since integrated signal will be read out every 10\arcsec . A significant portion (but not all) of the sky could be surveyed. In this paper we offer a preliminary analysis of an additional simultaneous all sky survey using these 2 detector channels. The specifications of the IRC all sky survey mode are shown in Table~\ref{IRC}.

The ASTRO-F all sky survey will be far superior to that of the IRAS survey. Optimally, ASTRO-F will have almost 2 orders of magnitude better sensitivity at 100$\umu$m. The detection limits for pointing observations in the near-mid-infrared are expected to be $\sim 1-150\umu Jy$ and 10-100mJy in the far-infrared (c.f. $\sim$0.5-1.5Jy for IRAS at 12, 25 ,60$\umu$m \& 100$\umu$m respectively ~\cite{soifer87}). ASTRO-F will also offer 4 times the resolution over 10 times the detector area compared to the ISOCAM instrument on ISO. Spatial resolution with IRAS was limited to 2-5arcmin. ASTRO-F will attain spatial resolutions of 50-70arcsec at wavelengths of 50-200$\umu$m respectively. Fig.~\ref{sensitivity} plots the sensitivities expected for the FIS All Sky Survey in table~\ref{FIS} against the SEDs of two sources from the the Canada-U.K. Deep SCUBA Survey (CUDSS, Lilly et al.~\shortcite{lilly99}, SED fits from Takagi et al~\shortcite{takagi02}): CUDSS10A and CUDSS03A, which have redshifts of 0.55 \& $\sim$1.6 and 850$\umu$m fluxes of 4.8mJy \& 3.4mJy respectively. Both of these sources would be detected in the All Sky Survey (CUDSS03A in only the 2 most sensitive FIS bands) and both sources would be easily detectable in all the 9 bands of the IRC instrument in pointing mode (see Pearson et al.~\shortcite{cpp01a} for IRC sensitivities in pointing mode).

NASA's SIRTF mission \cite{fazio98}, due for launch in early 2003, offers similar sensitivities to ASTRO-F. However these two missions should be seen as complementary rather than rivalling each other. ASTRO-F  will be able to reach such detection limits over a wider area (10-20sq. deg. due to the larger FOV, c.f., the SIRTF-SWIRE survey~\cite{lonsdale01}) in the mid-infrared and the entire sky at far-infrared wavelengths, providing the first All Sky Survey at 170$\umu$m. ASTRO-F will be sensitive to larger scales at lower redshifts while SIRTF while conversely be sensitive to smaller scales at higher redshifts. ASTRO-F is a surveyor, while SIRTF is an observatory (a similar comparison may be made between the IRAS and ISO missions). At 2.5-5 years, SIRTF has a longer projected lifetime than ASTRO-F, although it has no capability at near-infrared wavelengths ($<5\umu m$). ASTRO-F has this added capability and will in fact be able to continue to use this instrument even after liquid Helium exhaustion. 

At present, the ASTRO-F mission can be divided into 4 distinct phases.

\begin{enumerate}
  \item {\it PHASE 0 - Performance Verification} - $\sim$60 days - Initial check out and calibration of telescope and instruments. A mini-survey may also be attempted.
  \item {\it PHASE 1 - All Sky Survey} -  $\sim$180 days - ASTRO-F will scan the entire ecliptic longitude in survey mode. The All Sky Survey will have first priority although there may be as many as 1500 independent IRC pointings and $\sim$300 FIS pointings planned.
  \item {\it PHASE 2 - Pointing Observations} -  $\sim$300 days - Any supplemental survey observations will be performed. The rest of this phase will be dedicated to pointing observations with the IRC and FIS instruments. This phase will last until Helium exhaustion during which $\sim$6000 pointings are expected.
  \item {\it PHASE 3 - Helium Boil Off} - $>$365 days? - Even after the liquid Helium has been exhausted the mechanical coolers will keep the temperature low enough to enable the continuation of observations using the IRC-NIR detector. In this phase there is hope to be as many as 10500 pointings.
\end{enumerate}

\begin{figure}
\centering
\centerline{
\psfig{figure=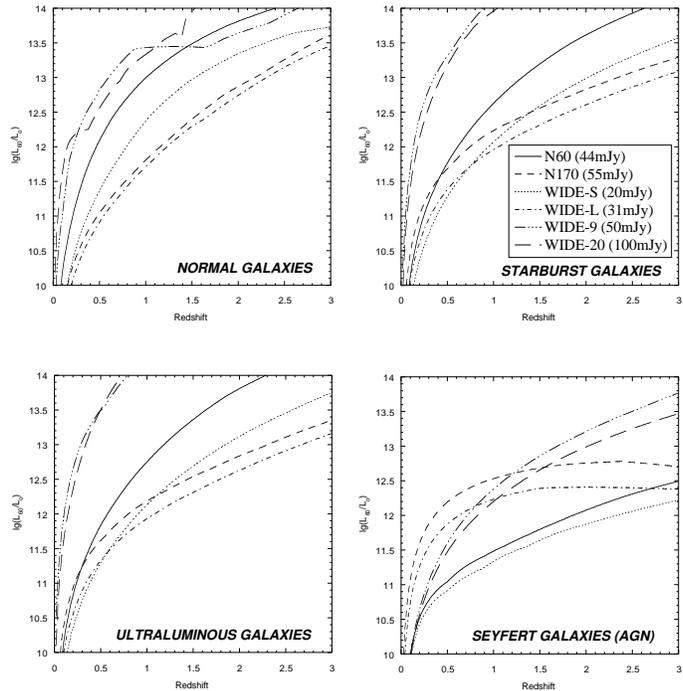,width=9cm}
}
\caption{Visibility of ASTRO-F All Sky Survey sources. The 60$\umu$m luminosity that a source must have, to be included in the All Sky Survey of sensitivity, S(mJy) at redshift, z. Results are plotted for the 4 far-IR bands (N60, N170, WIDE-S, WIDE-L) and proposed WIDE-9$\umu$m and WIDE-20$\umu$m mid-IR bands for the 4 components in the galaxy evolution model.
\label{lz}}
\end{figure}

\bigskip

\section{Model Parameters}\label{sec:model}

	The galaxy evolution models of Pearson~\shortcite{cpp01b} (hereafter CPP) are used to estimate the source counts, the number of sources detected and other observables such as number-redshift distributions and expected galaxy colours in the ASTRO-F All Sky Survey.
The CPP model is an improved extension of the original model of Pearson and Rowan-Robinson model ~\shortcite{cpp96} that provided an extremely good fit to IRAS data. The model incorporates a 4 component parameterization segregated by IRAS colours of galaxies ~\cite{mrr89} consisting of normal, starburst, ultraluminous galaxies ~\cite{sand96}, defined at 60$\umu$m and an AGN population defined at 12$\umu$m. The  cool (high S(100$\umu$m)/S(60$\umu$m)) and warm component (low S(100$\umu$m)/S(60$\umu$m)) 60$\umu$m luminosity functions of Saunders et al. ~\shortcite{saun90} are used to represent the normal, starburst \& ULIG galaxies respectively, the ULIGs comprising the high luminosity tail of the warm luminosity function ($L_{60\mu m} > \sim 10^{12} L\sun $). The AGN population utilizes the 12$\umu$m luminosity function of Lawrence et al. ~\shortcite{law86} modelled on the sample of Rush et al.~\shortcite{rush93} (This IRAS selected sample is consistent with the ISO subsample of Spinoglio et al.~\shortcite{spinoglio02}).

Spectral energy distributions are required both for the K-corrections of sources and to transform the luminosity functions from their rest wavelength to the observation wavelength (i.e. by a factor $L(\lambda _{obs})/L(\lambda _{LF} )$ where $\lambda _{LF}$ is the wavelength at which the luminosity function is defined and $\lambda _{obs} $ is the wavelength of observation. These model spectra templates are taken from the new radiative transfer models of Efstathiou, Rowan-Robinson \& Siebenmorgen ~\shortcite{esf00} and Efstathiou \& Rowan-Robinson ~\shortcite{esf02} for the starburst \& Ultraluminous galaxy components and the normal galaxy component respectively. The AGN spectral template is represented by the dust torus model of Rowan-Robinson ~\shortcite{mrr95}.

The CPP model assumes that the starburst, ULIG and AGN components evolve with cosmic time. The normal galaxy population does not evolve. 

The starburst and AGN components undergo positive luminosity evolution of the form $L(z)=L(0)(1+z)^{3.2}$ to a redshift of $z=2.5$ ~\cite{cpp96} which provides a good fit to the source counts, at least at brighter fluxes from the radio to the X-ray regime (e.g. Oliver et al.~\shortcite{oliver02}, Serjeant et al. ~\shortcite{serjeant00}, Pearson et al.~\shortcite{cpp97}, Pearson \& Rowan-Robinson~\shortcite{cpp96}, Benn et al.~\shortcite{benn93}, Boyle et al.~\shortcite{boyl88}). 

However, the subsequent strong evolution evident in the galaxy population at FIR wavelengths as detected by ISOPHOT on ISO (Dole et al.~\shortcite{dole00}, Efstathiou et al.~\shortcite{esf00a}), at MIR wavelengths with ISOCAM (especially at 15$\umu$m where the source counts span 2 orders of magnitude in flux (Oliver et al.~\shortcite{oliver97}, Flores et al.~\shortcite{flores99a}~\shortcite{flores99b}, Altieri et al.~\shortcite{altieri99}, Aussel et al.~\shortcite{aussel99}, Gruppioni et al.~\shortcite{gruppioni99}, Biviano et al.~\shortcite{biviano00}, Elbaz~\shortcite{elbaz00}, Serjeant et al.~\shortcite{serjeant00}) and in the sub-mm with SCUBA on the JCMT  (Hughes et al.~\shortcite{hugh98}, Smail et al.~\shortcite{smail97}, Scott et al.~\shortcite{scott00}) has introduced the need for some extreme evolutionary scenarios in galaxy evolution models.

Therefore, the ULIG population is assumed to undergo extreme evolution in both density and luminosity (see Pearson~\shortcite{cpp01b} for details of these evolutionary models). The density evolution is of the form $D(z) = 1 + g  \exp{[-{(z-z_{p})^{2}\over{2\sigma ^2}}]}$ with $g=250$, $z_{p}=0.8$, $\sigma =0.2$ to $z_{p}$ and decaying quickly towards higher redshifts. The luminosity evolution is of the form $L(z) = 1+ k \exp{[-{(z-z_{p})^{2}\over{2\sigma ^2}}]}$ with $k=40$, $z_{p}=2.5$, $\sigma =0.58$ to $z_{p}$ and decaying quickly to higher redshifts. In this scenario the luminosity evolution corresponds to the accretion of matter onto a core initiating star-formation at high redshifts. The density evolution corresponds to major mergers at a later epoch. Locally, ULIGs have a space density comparable to optically selected QSOs ($\sim 0.001per sq.deg.$), although analysis of the IRAS Faint Source Catalogue leads to the conclusion that they should be much more numerous at higher redshift ~\cite{kim98}. Moreover both the ISO 15$\umu$m differential source counts ~\cite{elbaz99} and the dust enshrouded star formation rate ~\cite{chary01}, ~\cite{elbaz02} exhibit a peak due sources at z$\sim$0.8. The evolutionary models of CPP provide an excellent fit to the source counts, number-redshift distributions and the integrated background radiation from sub-mm to NIR wavelengths (Pearson~\shortcite{cpp01b} and references there in).  

Although the FIR SED of galaxies is relatively featureless, the spectral templates are convolved with each individual FIS filter bandwidth to achieve smoothed model spectral templates for all components in all ASTRO-F FIS wavebands. The SEDs are also convolved with the 2 MIR wide bands. This convolution is more significant as unidentified infrared bands (UIBs probably due to PAH features in the galaxy SEDs ~\cite{pug89},~\cite{lu96},~\cite{boul96},~\cite{vig96}) may significantly affect any MIR observations ~\cite{xu98}~\cite{xu00}~\cite{cpp01a}. Indeed, the equivalent widths of these unidentified bands can be as much as 10$\umu$m, comparable to the ASTRO-F IRC-MIR wide filter band passes of S9W(6-11.5$\umu$m) and L20W(14-26$\umu$m) (see Table~\ref{IRC}).

Following growing consensus between observation and theory (e.g. Balbi et al.~\shortcite{balbi00}, Freedman et al.~\shortcite{freedman01}, Kravtsov et al.~\shortcite{kravtsov98}), the model assumes $H_o = 72kms^{-1}Mpc^{-1}, \Omega=0.3, \Lambda=0.7$ unless otherwise explicitly stated.

\bigskip

\section{Survey Predictions}\label{sec:survey}

	The IRAS All Sky Survey probed out to redshifts of $\approx$0.03-0.2, ASTRO-F is hoped to extend this range out to redshift $\geq$1. In order to assess the visibility of sources in the ASTRO-F All Sky Survey, we plot in fig.~\ref{lz} the luminosity-redshift distributions for each model galaxy component at each all sky survey wavelength. In essence fig.~\ref{lz} plots the rest 60$\umu$m luminosity that a source must have if it is to be detected in the survey in a given wavelength band with limiting flux {\it S}. We see that the normal galaxy population is quickly shifted beyond $lg(L_{60} /L_{\sun} )=11.5$ (the arbitrary limit for the normal galaxy population in CPP) at redshifts of $\sim$0.3 \& 0.8 in the N60 \& N170 bands respectively, i.e. there will be no high redshift (z$>$1) normal galaxies in the sample. However, the starburst population (starbursts, LIGs, ULIGs) enjoy a stronger {\it negative} K-correction in the longer wavelength bands as the band pass climbs the Rayleigh-Jeans slope towards the dust peak around 170-60$\umu$m in the observed frame. Thus a $L_{60} =10^{12} L_{\sun} $ULIG  (corresponding to a total FIR luminosity of $\approx 2.5$x$10^{12}L_{\sun}$~\cite{helou84}) should be detectable out to redshifts of z$\sim$1 in the long wavelength bands. Similarly the AGN have an almost flat luminosity curve out to redshifts of 2.5-3 as the wavelength band pass slowly move up the long wavelength SED towards the peak at around 30$\umu$m. Figure~\ref{lz} also emphasises the constraints placed on the mid-IR All Sky Survey by their relatively low sensitivity. At mid-IR wavelengths almost all the sources would be expected at z$\ll$1 except for the AGN that have high luminosities and are also enjoying a sampling wavelength closer to the peak in their SED. In summary, from fig.~\ref{lz} we may broadly expect 4 regimes of increasing redshift encompassing the normal, starburst, ULIG and AGN populations respectively.

\begin{table*}
\caption{Estimated confusion limits due to point sources for ASTRO-F All Sky Survey bands}
\begin{tabular}{@{}lccccc}
Band & Flux & Confusion Source Density & \multicolumn{3}{c}{Confusion Limit (mJy)}  \\
 & 5$\sigma$ mJy &  (Number of Sources) & $\Omega =0.3, \Lambda = 0.7$ & $\Omega =0.3, \Lambda = 0$& $\Omega =1, \Lambda = 0$ \\
\hline
$N60$     &$44$  & $840$ & $9$ & $9$ & $9$ \\
$N170$    &$55$  & $100$ & $83$ & $86$ & $97$ \\ 
$WIDE-S$  &$20$  & $540$ & $20$ & $20$ & $22$ \\
$WIDE-L$  &$31$  & $120$ & $79$ & $82$ & $91$ \\  
$WIDE-9$  &$50$  & $37300$ & $0.007$ & $0.007$ & $0.004$ \\ 
$WIDE-20$ &$100$ & $7560$ & $0.16$ & $0.16$ & $0.12$ \\ 
\hline\noalign{\smallskip}
\noalign{\smallskip}
\end{tabular}
\label{confuse}
\end{table*}

\begin{figure*}
\centering
\centerline{
\psfig{figure=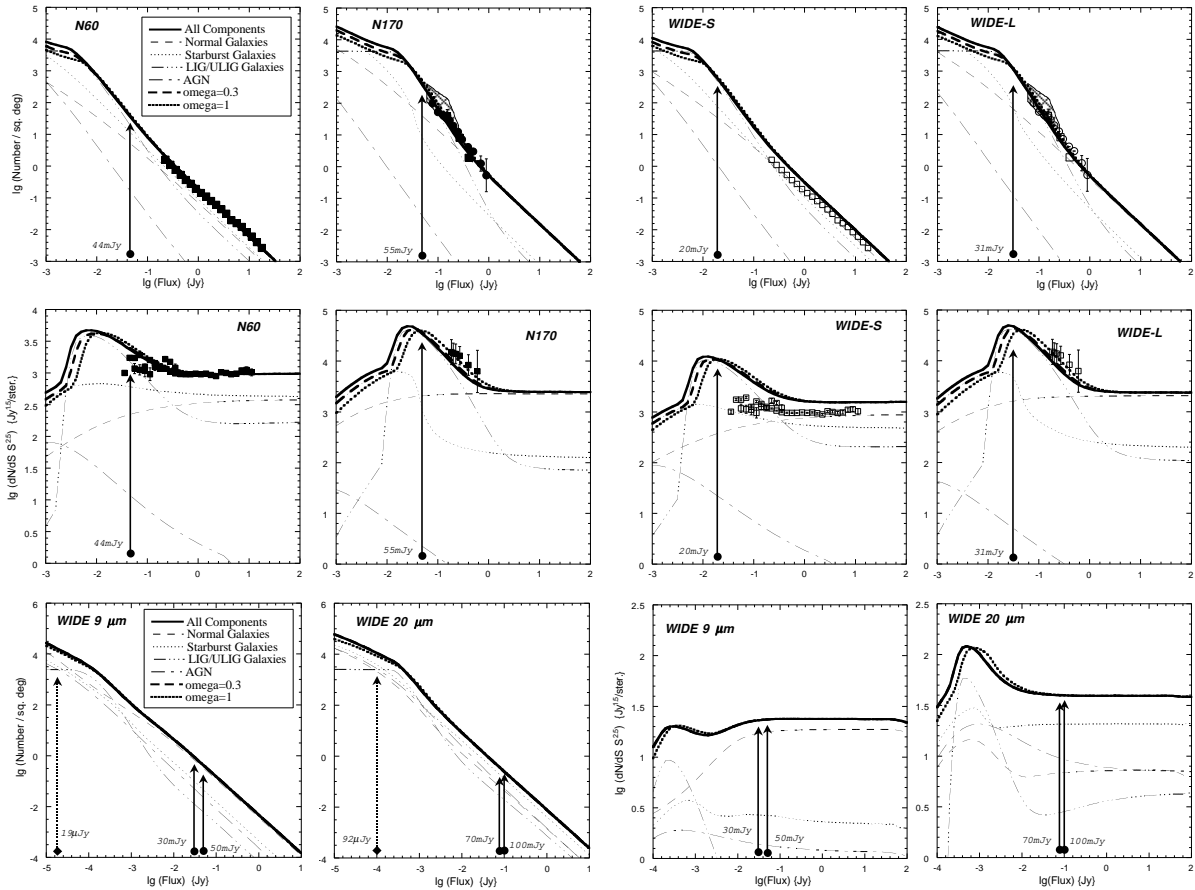,width=170mm}
}
\caption{Predicted integral(top row) and normalized differential (middle row) source counts for the ASTRO-F All Sky Survey in the 4 far-IR bands (N60, N170, WIDE-S, WIDE-L) compared with expected sensitivities in mJy. Observed counts shown for N60 band are from IRAS data at 60$\umu$m - Lonsdale et al. (1990), Hacking \& Houck (1987), Rowan-Robinson et al. (1990), Saunders (1990), Gregorich et al. (1995). Observed counts shown for N170 band are from ISO data at 170$\umu$m - Kawara et al. (1998), Puget et al. (1999), Matsuhara et al. (2000) (fluctuation analysis - {\it dotted box\,}), Dole et al. (2001). The observed data at 60$\umu$m and 170$\umu$m are also overplotted as unfilled symbols for comparison, on the counts corresponding to the WIDE-S and WIDE-L bands. The bottom row shows the predicted source counts at mid-IR wavelengths in the proposed WIDE-9$\umu$m and WIDE-20$\umu$m All Sky mid-IR survey bands. The sensitivities shown correspond to the nominal survey mode (highest values), enhanced survey sensitivity obtained by a lower data sampling rate in mJy ({\it solid arrows\,}) and the 5$\sigma$ pointing sensitivity of the IRC instrument in $\umu$Jy ({\it dotted arrow\,}). 
\label{counts}}
\end{figure*}

\begin{figure*}
\centering
\begin{minipage}[c]{170mm}
\psfig{angle=90, figure=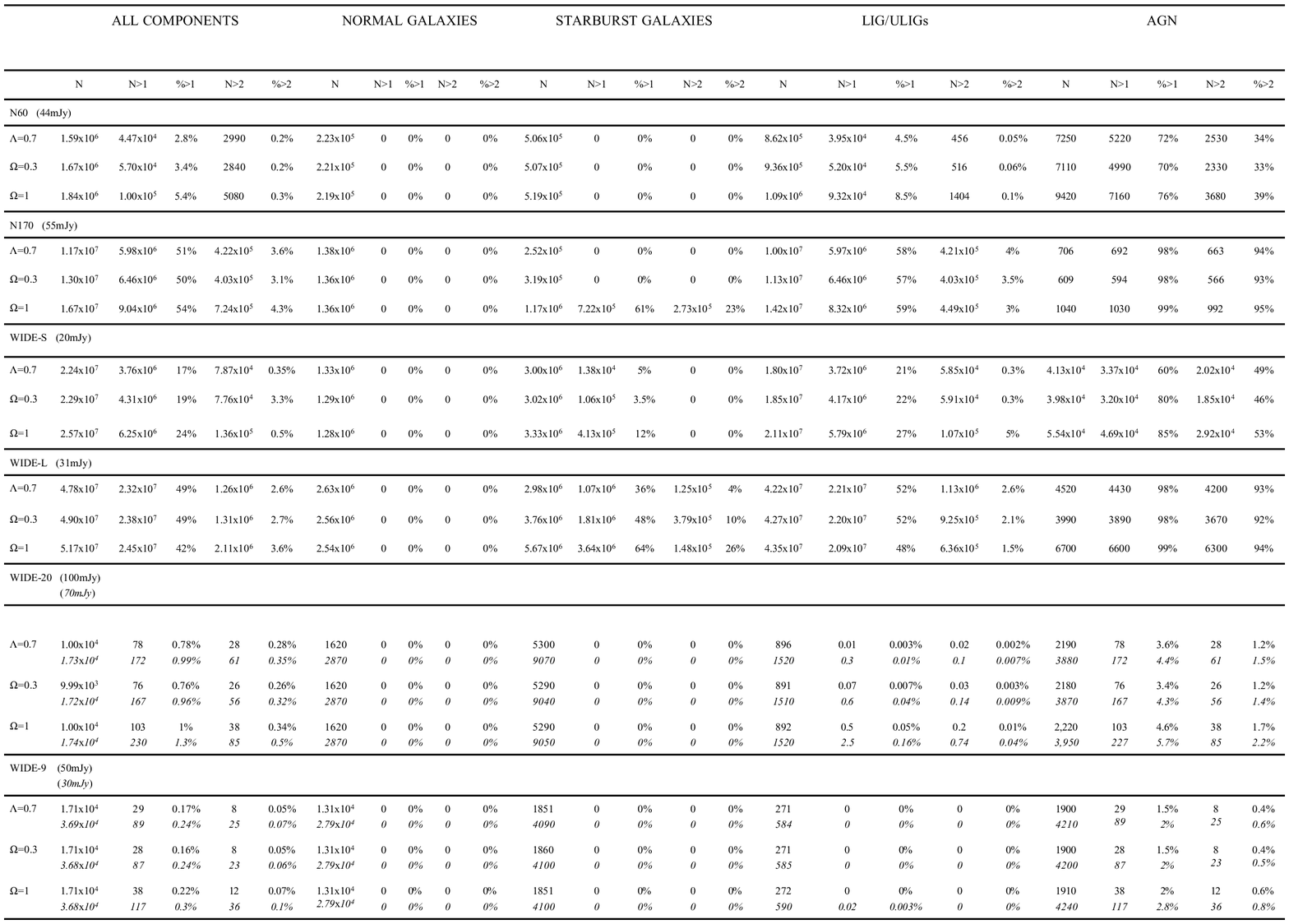, width=165mm}
\end{minipage}
\begin{minipage}[c]{5mm}
\rotcaption{ASTRO-F All Sky Survey model predictions at expected 5$\sigma$ sensitivities for 3 cosmological models a flat $\Omega =0.3,\Lambda=0.7$ universe, an open $\Omega =0.3,\Lambda=0$ universe and a flat $\Omega =1,\Lambda=0$ universe. Predicted number of sources segregated by galaxy type for total numbers, number higher than a redshift of 1, number higher than a redshift of 2 and percentage fractions. Numbers in italics for the MIR bands are the expectations assuming a higher sensitivity due to a lower sampling rate.
\label{numbers}}
\end{minipage}
\end{figure*}

\subsection{Source Counts}

	The ASTRO-F All Sky Survey will provide a huge database for the statistical study of galaxies. The zeroth order moment of this statistical data base (for a complete sample of sources) is simply the number counts of detected sources. Integral galaxy number counts as a function of flux will indeed form the {\it front line} of the results from the ASTRO-F mission.  Fig.~\ref{counts} shows the predicted source counts calculated from the model explained in section~\ref{sec:model}. Numbers for the All sky survey are calculated using 3 different cosmologies assuming a Hubble constant of $72kms^{-1}Mpc^{-1}$, a flat  $\Omega =0.3,\Lambda=0.7$ universe, an open $\Omega =0.3,\Lambda=0$ universe and a flat $\Omega =1,\Lambda=0$ universe (note, changing the value of the Hubble constant will have no effect on the source counts).

 The source counts in the N60 and N170 narrow bands can be compared directly with observations by IRAS and ISO respectively whilst counts in the wide far-IR bands will probe down to and below the confusion limit due to point sources. In general, the ultimate sensitivity for any far-IR telescope at high galactic latitude will be determined by one of 3 contesting noise values. The instrument noise, the noise due to diffuse IR cirrus emission ~\cite{low84} which is spatially dependent or the confusion limit due to point sources ~\cite{helou90}.  In table ~\ref{confuse} we calculate the limiting flux sensitivity corresponding to the confusion limit due to point sources for each of the ASTRO-F All Sky Survey Bands for 3 different cosmological world models. We assume the classical confusion criteria of a source density of 1 source per $\sim$40 beams~\cite{condon74},~\cite{hogg01} of the observing instrument, where the beam diameter is given by $d=1.2 \lambda /D$, where $D$ is the telescope diameter (67cm). Here we calculate the source confusion assuming a beam diameter equivalent to the FWHM of the Airy disc, $d=1.2 \lambda /D$ such that the beam diameter at 170$\umu$m is $\sim$1arcmin corresponding to a source confusion density of $\approx$100 sources (where the subsequently calculated limiting flux will of course be model dependent). From table~\ref{confuse} we see that for the shorter FIS bands and the MIR bands the confusion limit due to point sources is not really dependent on the world model assumed and is on or well below the expected sensitivity capabilities of the instruments. For the longer wavelength bands (N170, WIDE-L) we would expect that the galaxy counts will be quite severely restricted by the source confusion due to the extremely steep slope of the source counts expected ($\sim$3.3 compared to 2.5 expected for a non-evolving population in a Euclidean universe (Kawara et al.~\shortcite{kawara98}, Dole et al.~\shortcite{dole01})). Indeed, observations at 175$\umu$m from the ISO FIRBACK survey~\cite{lagache00b} and Japanese Lockman Hole survey~\cite{mat00} place even more severe limits on the source confusion level suggesting that the confusion due to point sources would become severe at flux levels of $\sim$45mJy. Although SIRTF boasts a larger diameter mirror (85cm) than ASTRO-F and also has much better sensitivity at 70$\umu$m (2.75mJy) and 160$\umu$m (17.5mJy), it too will feel the pinch of the source confusion. Estimated limits for SIRTF are 11mJy at 70$\umu$m and 60mJy at 160$\umu$m. It should be noted that various innovative  fluctuation analysis methods can be used to push below these limits (e.g. Matsuhara et al.~\shortcite{mat00}).

 The source counts in fig.~\ref{counts} are plotted by galaxy component and tabulated in fig.~\ref{numbers}. The source counts are rapidly increasing down to fluxes of 10-20mJy where they begin to flatten off. Approximately, 1.6 million sources will be detected in the N60 band compared to more than 10 million in the N170 band. The N170 band samples higher up the source count slope (closer to the turn over) and is more sensitive to higher redshift galaxies due to the negative K-corrections (see fig.~\ref{lz}). The dust hump at the peak of the emission spectrum at 100$\umu$m and 60$\umu$m for normal and starburst galaxies passes through the N170 band at redshifts of 0.7 and  1.8 respectively where as the emission in the N60 band will already be decreasing. In the wide bands, approximately 22 million and 47 million galaxies may be detected in the WIDE-S and WIDE-L bands respectively. The vast majority of sources detected in all bands will be luminous and ultraluminous galaxies. Expected numbers range from 1-40 million from N60 - WIDE-L respectively although the exact numbers will be dependent on the assumed evolutionary scenario. The longer wavelength far-IR bands (N170/WIDE-L) will be particularly sensitive to these galaxies. AGN will be preferentially detected in the short wavelength FIR bands (N60/WIDE-S) as the AGN emission peaks at $\sim$25$\umu$m meaning that the longer wavelengths are still climbing the the Rayleigh-Jeans slope of the SED out to redshifts of $\sim$6 (see fig.~\ref{lz}). Approximately 7000 AGN will be detected in the N60 band, with as many as 40,000 in the superior WIDE-S band. The N170 band will detect between 700-1000 while the WIDE-L band should see as many as 4500 (these numbers being dependent on the assumed cosmology). 

Comparison with the IRAS integral counts at 60$\umu$m (N60 \& WIDE-S bands) shows that the ASTRO-F survey will probe between 3-10 times deeper than the various IRAS surveys (0.5Jy for IRAS-PSC  and 120mJy for the VFSS Gregorich et al.~\shortcite{gregor95}, Bertin et al. ~\shortcite{bertin97}). Moreover, a comparison with the IRAS 60$\umu$m differential source counts (Hacking \& Houck \shortcite{hacking87}, Rowan-Robinson et al. \shortcite{mrr90}, Saunders \shortcite{saun90p}, Gregorich et al. \shortcite{gregor95}) emphasizes that ASTRO-F will be reaching very interesting flux levels where some discrimination between competing evolutionary scenarios will be possible. At the limit of the IRAS-PSC the counts are Euclidean (i.e. flat in the normalized differential counts - the middle panel of figure~\ref{counts}), the IRAS-VFSS survey extended this flux limit down to $\approx$120mJy and suggested an upturn in the differential source counts at $\sim$200-300mJy similar to that observed in the differential counts at radio wavelengths due to the emergence of a radio sub-mJy population of starburst galaxies ~\cite{condon84}. Kim \& Saunders~\shortcite{kim98} re-analysed the IRAS-FSS data concluding that there was evidence of much stronger evolution at higher redshifts to z$\sim$1. The nature of this evolution cannot be constrained by the IRAS source counts at 60$\umu$m with both luminosity evolution or density evolution being equally acceptable (Pearson~\shortcite{cpp01b}, Oliver et al.~\shortcite{oliver92}). ASTRO-F will be able to detect and quantify the upturn in the differential counts at fluxes $<$200mJy.

ASTRO-F will be the first ever All Sky Survey at 170$\umu$m.  Since the peak of the energy output of dust enshrouded star formation lies in the FIR at $\sim$60$\umu$m the N170 and WIDE-L bands will be particularly sensitive to high redshift (z$>$1) galaxies as the peak of the SED emission is redshifted to 170$\umu$m at z$\sim$1.8 producing the phenonemon where galaxies are brighter at higher z than lower z as made apparent in fig.~\ref{lz}, \cite{franceschini91},\cite{cpp96}. To date the largest survey at 170$\umu$m was carried out by the FIRBACK team covering 4sq.deg. down to 180mJy ~\cite{dole01}. The ASTRO-F All Sky Survey will be aptly placed below the limit of the FIRBACK survey in both the N170 and WIDE-L bands. Although the WIDE-L band will almost certainly be source confused, valuable work can still be carried out on and below the confusion limit (c.f. the fluctuation analysis of ISO 170$\umu$m source counts \cite{mat00}). The sensitivity of the FIR bands will also be good enough to provide  discrimination between different evolutionary scenarios in the differential counts. The FIRBACK differential counts have already shown a steep non-Euclidean slope of 3.3$\pm$0.6 at fluxes less than 500mJy (c.f. a Euclidean slope of 2.5). ASTRO-F should detect any turnover in the differential source counts.

The lower panels of figure ~\ref{counts} shows the predictions for the possible additional survey in the 2 wide MIR wavebands at 9$\umu$m and 20$\umu$m. In this case, the survey sensitivities will be somewhat lower than compared to the FIR bands. This is due to constraints on the data transmission rate of the ASTRO-F satellite. Nominal sensitivities are currently 50mJy \& 100mJy at 9$\umu$m \& 20$\umu$m respectively with some possibility of increasing this to 30mJy and 70mJy respectively by using a lower sampling rate. Predicted numbers in the WIDE-9$\umu$m band are approximately 17,000 sources in total, mostly at local redshifts with approximately 2000 starburst galaxies and 2000 AGN. Results from the longer WIDE-20$\umu$m band are similar with approximately 10,000 sources in total of which around half would be star forming galaxies. Note that in the case of IRAS only 893 galaxies were obtained from the IRAS-FSS ($|b|>$25, $>$0.22Jy at 12$\umu$m \cite{moshir91},~\cite{rush93}). Furthermore, around 1000 sources detected by ISO were used to construct the ISO number counts at 15$\umu$m (Elbaz et al.~\shortcite{elbaz99}, Gruppioni et al.~\shortcite{gruppioni99}, Serjeant et al.~\shortcite{serjeant00}). ASTRO-F will obtain an order of magnitude larger sample than the IRAS sample and the ISO 15$\umu$m samples. Thus the local luminosity function of IR galaxies can be accurately derived since all galaxies at 9$\umu$m must be detected by the FIS survey.

ASTRO-F will also be an ideal tool for the discovery of rare objects such as hyperluminous ($L_{60\umu m} > \sim 10^{13} L_{\sun} $, ~\cite{mrr00a}) and primeval galaxies, c.f. IRAS F10214+4724 discovered at the limit of the IRAS-FSS with $S_{60} $=0.2Jy, at z=2.286 and far-IR luminosity $\sim 10^{14} L_{\sun}$ ~\cite{mrr91b}. Totani \& Takeuchi ~\shortcite{totani02} have suggested that starbursting primordial ellipticals may have higher dust temperatures (40-80K) than previously assumed. Such star bursting primordial ellipticals may also be observed in the ASTRO-F All Sky Survey. The best strategy for such detections are (i) Wide area shallow as opposed to narrow, deep surveys, since these objects are rare but bright with source densities of between 0.001-0.01per sq.deg. depending on the evolutionary scenario and cosmological world model. (ii) Far-IR wavelengths since the bulk of any emission (90-99$\%$) should emerge at these wavelengths due to the copious amounts of dust expected due to starformation. (iii) Multi-wavelength survey for IR colours for photometric redshifts, SED fitting and discrimination from other less luminous sources. An F10214 type hyperluminous galaxy with luminosity of order a few $10^{14} L{\sun} $ would be detectable out to redshifts $\sim$4 in the FIS bands and at z$>$1 in the IRC-MIR bands. Depending on the cosmology, we could expect of the order of 300 such sources over the entire sky detected in the WIDE-S and WIDE-L bands and between 40-100 in the WIDE-9 and WIDE-20 MIR bands respectively. Corresponding follow up and cross correlations with future mm wave instruments such as LMT (www.lmtgtm.org, e.g.  a large area survey of a few 1000sq.deg.) would be particularly sensitive,  will further constrain the detections of such objects since a detection at mm wavelengths with no corresponding ASTRO-F counterpart will automatically imply a high redshift (and therefore high luminosity) for any given source.

\begin{table*}
\caption{Expectation Redshift Values (Mean, Mode, Median) for the ASTRO-F All Sky Survey}
\begin{tabular}{@{}lcccccccccccccccc}
Band & Flux & \multicolumn{3}{c}{Total} & \multicolumn{3}{c}{Normal Galaxies} & \multicolumn{3}{c}{Starburst Galaxies}& \multicolumn{3}{c}{LIG/ULIG} & \multicolumn{3}{c}{AGN}  \\
 & 5$\sigma$ mJy &  mean & mode & med  &  mean & mode & med  &  mean & mode & med  &  mean & mode & med  &  mean & mode & med \\
\hline
$N60$     &$44$    & $0.33$ & $0.05$ & $0.29$ & $0.06$ & $0.05$ & $0.06$ & $0.17$ & $0.15$  & $0.16$ & $0.49$ & $0.35$ & $0.44$ & $1.56$ & $0.05$ & $1.54$ \\
$N170$    &$55$    & $0.96$ & $1.25$ & $0.99$ & $0.12$ & $0.05$ & $0.10$ & $0.20$ & $0.05$ & $0.18$ & $1.10$ & $1.25$ & $1.09$ & $3.44$ & $3.25$ & $3.49$\\
$WIDE-S$  &$20$    & $0.67$ & $0.55$ & $0.63$ & $0.09$ & $0.05$ & $0.08$ & $0.37$ & $0.25$ & $0.34$ & $0.76$ & $0.55$ & $0.69$ & $1.89$ & $0.05$ & $1.91$   \\
$WIDE-L$  &$31$    & $0.98$ & $1.15$ & $0.97$ & $0.14$ & $0.05$ & $0.13$ & $0.83$ & $0.25$ & $0.71$ & $1.04$ & $1.15$ & $1.02$ & $3.17$ & $2.95$ & $3.15$  \\
$WIDE-9$  &$50$    & $0.06$ & $0.05$ & $0.05$ & $0.05$ & $0.05$ & $0.05$ & $0.05$ & $0.05$ & $0.05$ & $0.06$ & $0.05$ & $0.05$ & $0.10$ & $0.05$ & $0.06$ \\
$WIDE-20$ &$100$   & $0.07$ & $0.05$ & $0.05$ & $0.05$ & $0.05$ & $0.05$ & $0.05$ & $0.05$ & $0.05$ & $0.07$ & $0.05$ & $0.06$ & $0.15$ & $0.05$ & $0.06$\\
\hline\noalign{\smallskip}
\hline
\noalign{\smallskip}
\end{tabular}
\label{mean}
\end{table*}

\begin{figure}
\centering
\centerline{
\psfig{figure=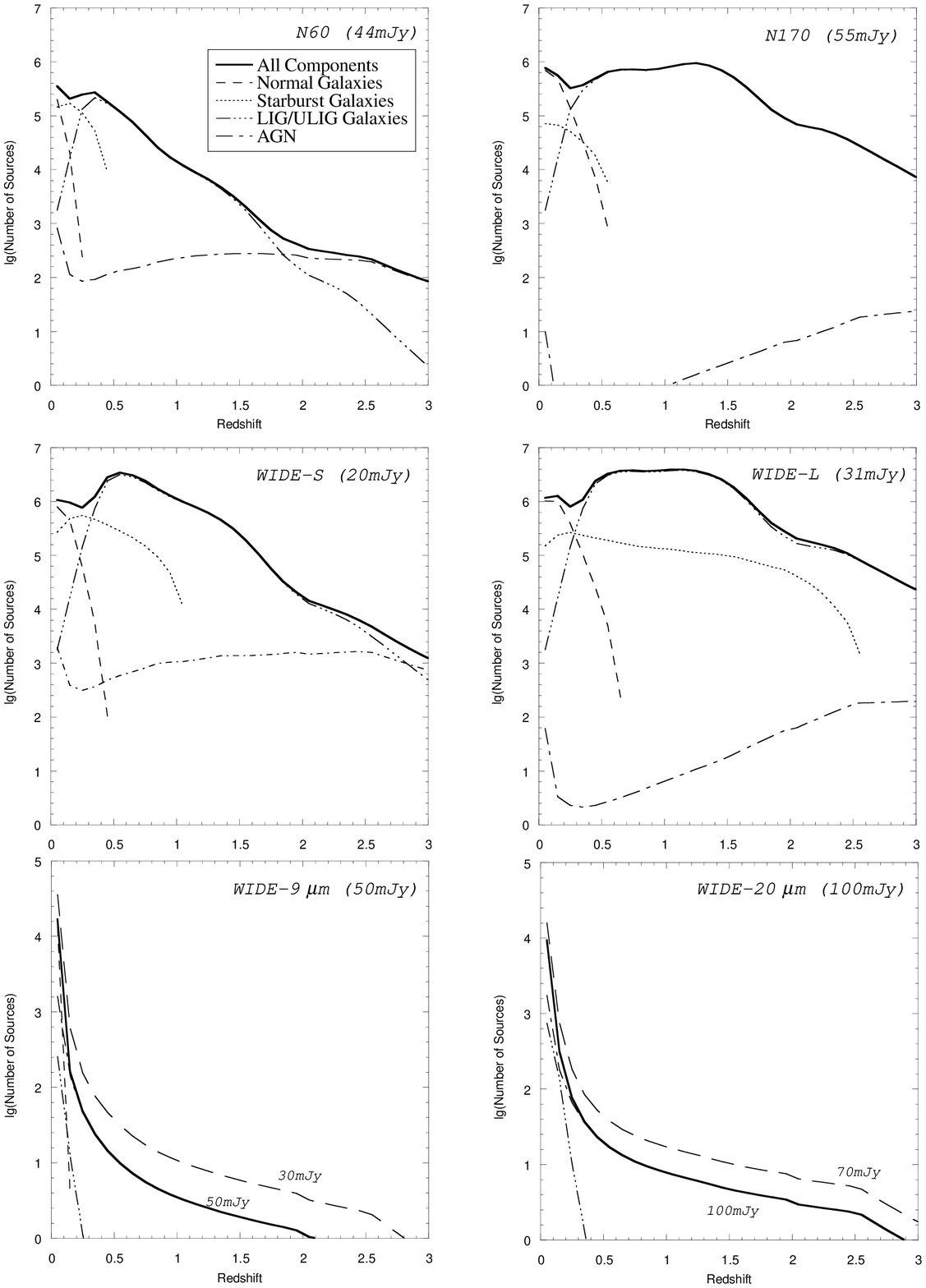,width=90mm}
}
\caption{\small Predicted Number-Redshift distributions for the ASTRO-F All Sky Survey in the 4 far-IR bands (N60, N170, WIDE-S, WIDE-L) and proposed WIDE-9$\umu$m and WIDE-20$\umu$m mid-IR bands assuming 5$\sigma$ survey sensitivities in mJy. Also shown for the 2 mid-IR bands are the total contribution assuming an enhanced survey sensitivity obtained by a lower data sampling rate({\it long dashed lines\,}). Vertical axis gives the total number of sources contained within a redshift bin of width $\delta$z=0.1. A $\Omega =0.3,\Lambda=0.7$ cosmology is assumed. 
\label{nz}}
\end{figure}

\begin{figure}
\centering
\centerline{
\psfig{figure=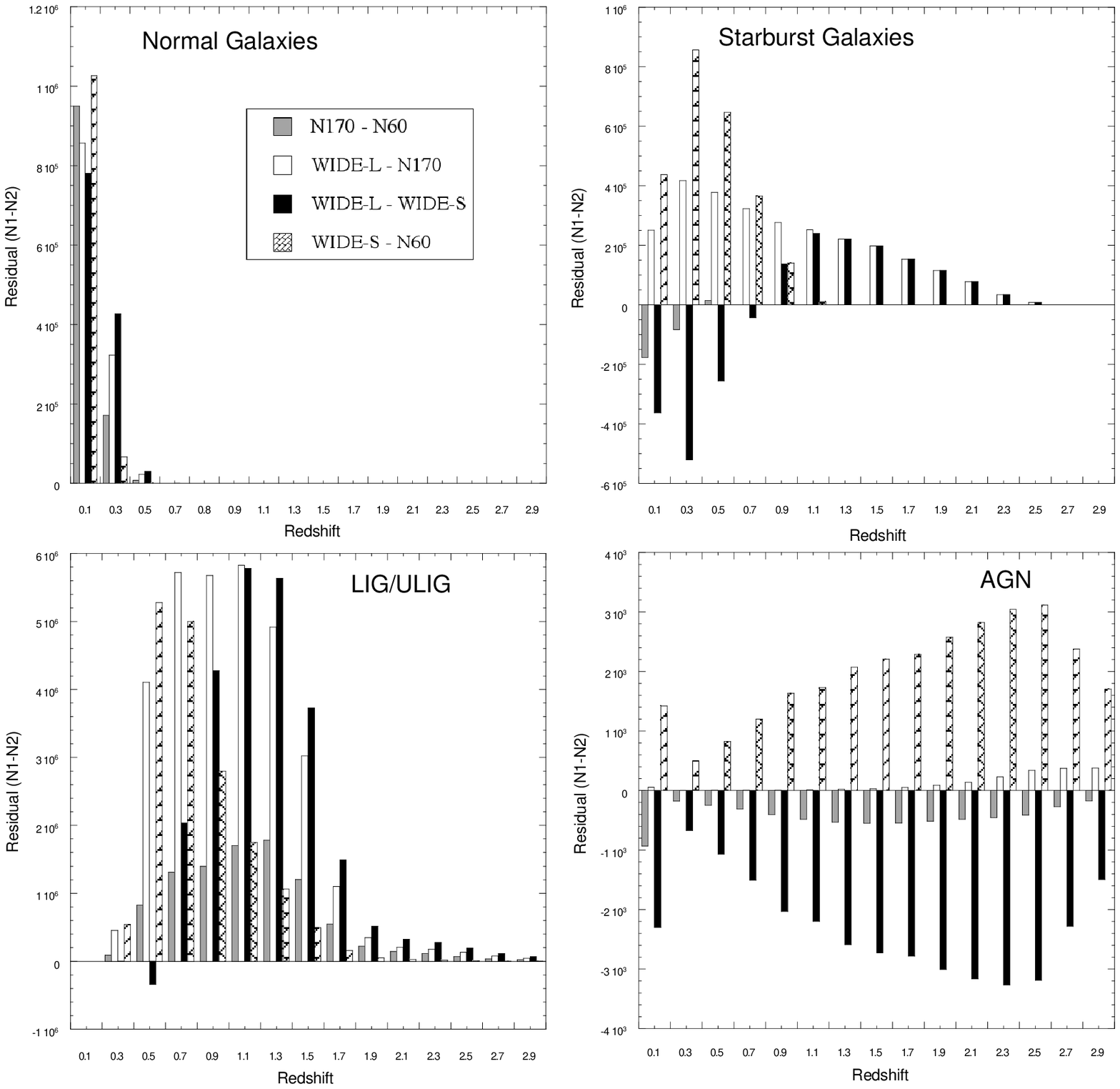,width=11cm}
}
\caption{The residual Number-Redshift distributions calcualted from the difference in the N-z distributions at 2 wavelengths for combinations given in the figure legend. Vertical axis gives residual number as $N( \lambda _{1})-N( \lambda _{2})$. A positive value indicating that $N( \lambda _{1}) > N( \lambda _{2})$ and a negative value indicating that $N( \lambda _{1}) < N( \lambda _{2})$. A value of zero indicates that all galaxies are seen in both wavelength bands.
\label{residual}}
\end{figure}

\subsection{Number-Redshift Distributions}

	Figure ~\ref{nz} shows the expected number redshift distributions for the ASTRO-F All Sky Survey in the 4 FIS bands along with predictions for the MIR All Sky Survey currently under consideration. Beyond a redshift of $\sim$0.3 the number redshift distributions in all the FIS bands are dominated by luminous/ultraluminous infrared galaxies consistent with the results of deep surveys with ISO (e.g. Elbaz et al.~\shortcite{elbaz02}, Dole et al.~\shortcite{dole01}). The narrow FIS bands (N60,N170) are only sensitive to normal and starburst galaxies out to redshifts of $\sim$0.5, the shortest band only detecting a few percent of these galaxies above a redshift of unity. The longer wavelength bands, due to the large K-corrections encountered (see fig.~\ref{lz}),  do better at detecting sources at z$>$1 with about half the total number of detections expected to be at redshifts higher than unity. The WIDE-L band (and to a lesser extent the WIDE-S band), due to their better sensitivity are also sensitive to starburst galaxies at z$\geq$1 with between 1/20-1/8 and 1/3-2/3 depending on the cosmology being found at z$>$1 in the WIDE-S and WIDE-L bands respectively. Due to the nature of the assumed AGN SED almost all AGN detected in the long wavelength bands will be at z$>$2 (30-50$\%$ for the short wavelength bands), thus providing a unique sample of far-infrared, high redshift AGN especially at 170$\umu$m where the low detection rate means that only a large area (all sky) survey such as the ASTRO-F survey can provide us with a statistically viable sample of such objects. In summary, the ASTRO-F survey will provide the database for a comprehensive redshift survey out to redshift $\sim$1 for normal and starburst galaxies and to redshift 2 and higher for ULIG and AGN sources thus bridging the dark gap between the IRAS observations in the local Universe and the optical and sub-mm studies in the high redshift regime.

To investigate any correlation between the Number-Redshift distributions further, we plot in fig.~\ref{residual}, the N-z Residual. That is, the difference between the N-z distributions at 2 wavelengths. Residual numbers are plotted as $N( \lambda _{1})-N( \lambda _{2})$. A positive value indicates that $N( \lambda _{1}) > N( \lambda _{2})$, a negative value that $N( \lambda _{1}) < N( \lambda _{2})$ and a  value of zero indicates that all galaxies are seen in both wavelength bands, i.e. that the N-z distributions are identical. We conclude that normal galaxies are preferentially detected in the long wavelength bands due to the fact that the peak of their SED is around 100$\umu$m  as are the LIG/ULIG sources due to the strong K-corrections incurred at redshifts up to $\sim$1.8. For starburst galaxies, at lower redshifts we are sampling the dust emission hump at $\sim$60$\umu$m, thus the sources are preferentially detected in the short wavelength bands, moving to higher redshift shifts the residuals in favour of the longer wavelength bands. For AGN, the shorter wavelength bands are preferred due to the flat nature of the SED from 30$\sim$3$\umu$m. The 60$\umu$m band climbs out of the Rayleigh-Jeans region of the SED at z$\approx$1 whereas the 170$\umu$m remains on the Rayleigh-Jeans slope until z$>$4.

 Note, the residuals corresponding to the wide-narrow band passes simply show the increase in detections due to the enhanced sensitivity of the wide band filters over the narrow band filters.

We find average redshifts, median(mean), of 0.29(0.34), 0.99(0.96), 0.63(0.67), 0.97(0.98) for the N60, N170, WIDE-S, WIDE-L bands respectively. By comparison the IRAS QDOT sample had a median redshift of$\sim$=0.03 ~\cite{mrr91a} and even the IRAS Very Faint Source Survey analysis  was limited to z$<$0.3~\cite{gregor95}. In Table~\ref{mean} a detailed analysis of the expected redshift values (mean, mode, median) is tabulated for all bands and all components for the the ASTRO-F All Sky Survey. Distinct populations sample distinct redshift regimes at factors of 2-10 deeper than the IRAS surveys.

The number-redshift distribution of the proposed MIR All Sky Survey at 9$\umu$m and 20$\umu$m shown in the bottom pair of panels in fig.~\ref{nz} is somewhat limited to redshifts $<$0.5 (z$<$1 in the sparse sampling rate scenario - see the figures in italics in fig.~\ref{numbers}). However, even this relatively shallow sample will enable a more accurate determination of the local 9$\umu$m and 20$\umu$m luminosity functions. Previous recent studies of the 12$\umu$m and 25$\umu$m luminosity functions with IRAS have been limited to z$<$0.3 at $S_{25} > 225mJy$ (Fang et al.~\shortcite{fang98}, Shupe et al.~\shortcite{shupe98}). Even with the limitations of the proposed all sky MIR survey (median redshift $\approx$0.05, the MIR luminosity function of galaxies will be accurately determined out to at least a factor of 2 higher in redshift and a factor of 2-5 better sensitivity in flux.

\begin{figure*}
\centering
\centerline{
\psfig{figure=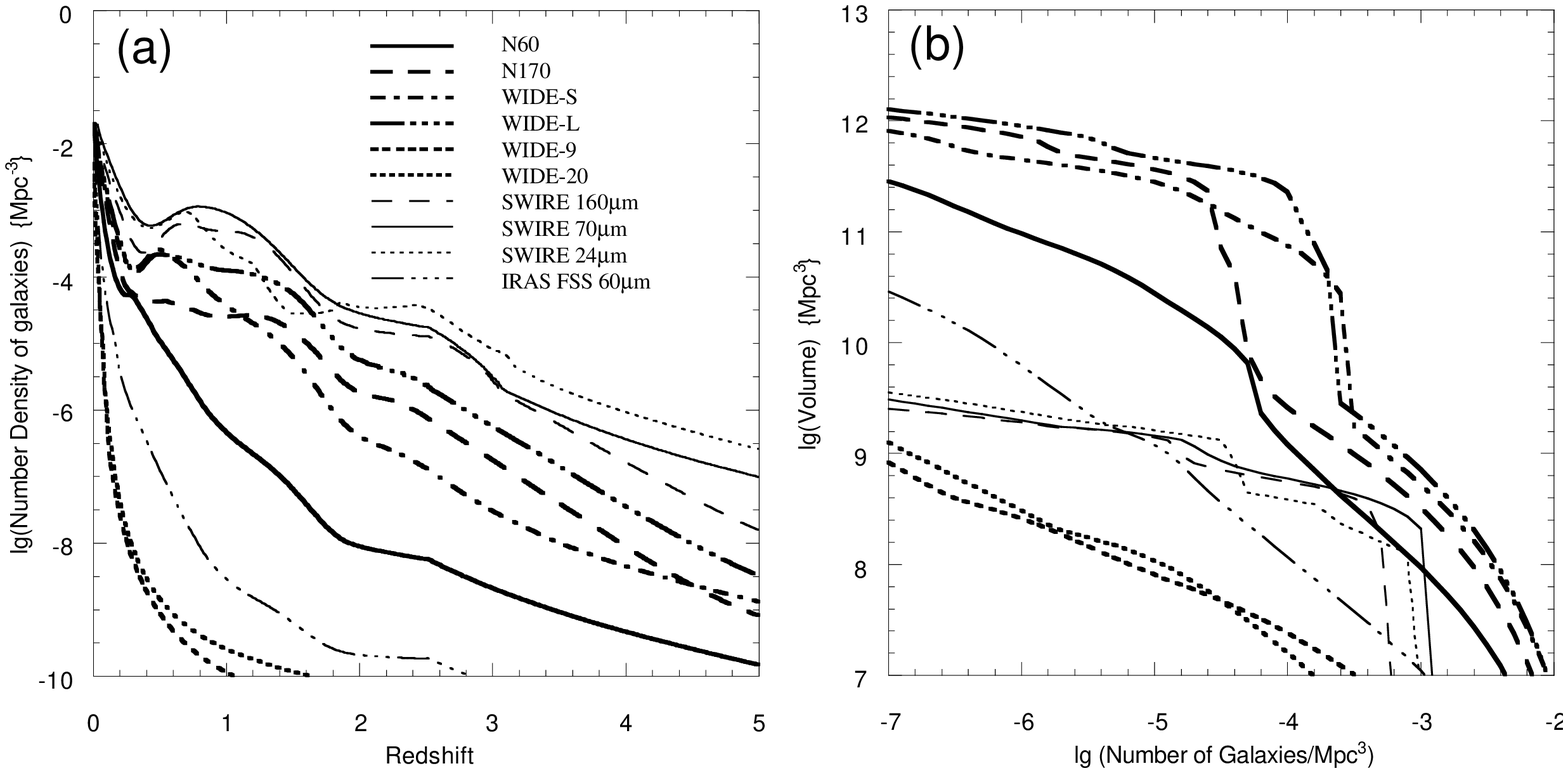,width=170mm}
}
\caption{The Density Function (a) and the Visibility Function (b) of the ASTRO-F All Sky Survey. Results are plotted for the 4 far-IR bands N60 (S=44mJy), N170 (S=55mJy), WIDE-S (S=20mJy), WIDE-L(S=31mJy) and proposed WIDE-9$\umu$m (S=50mJy) and WIDE-20$\umu$m (S=100mJy) mid-IR bands. Also shown for comparison are the predictions for the SIRTF-SWIRE survey covering 70sq.deg. at wavelengths of 160$\umu$m (S=17.5mJy), 70$\umu$m (S=2.75mJy), 24$\umu$m (S=0.45mJy) and the IRAS FSS all sky survey at 60$\umu$m (S=200mJy).  
\label{vis}}
\end{figure*}

\begin{figure*}
\centering
\centerline{
\psfig{figure=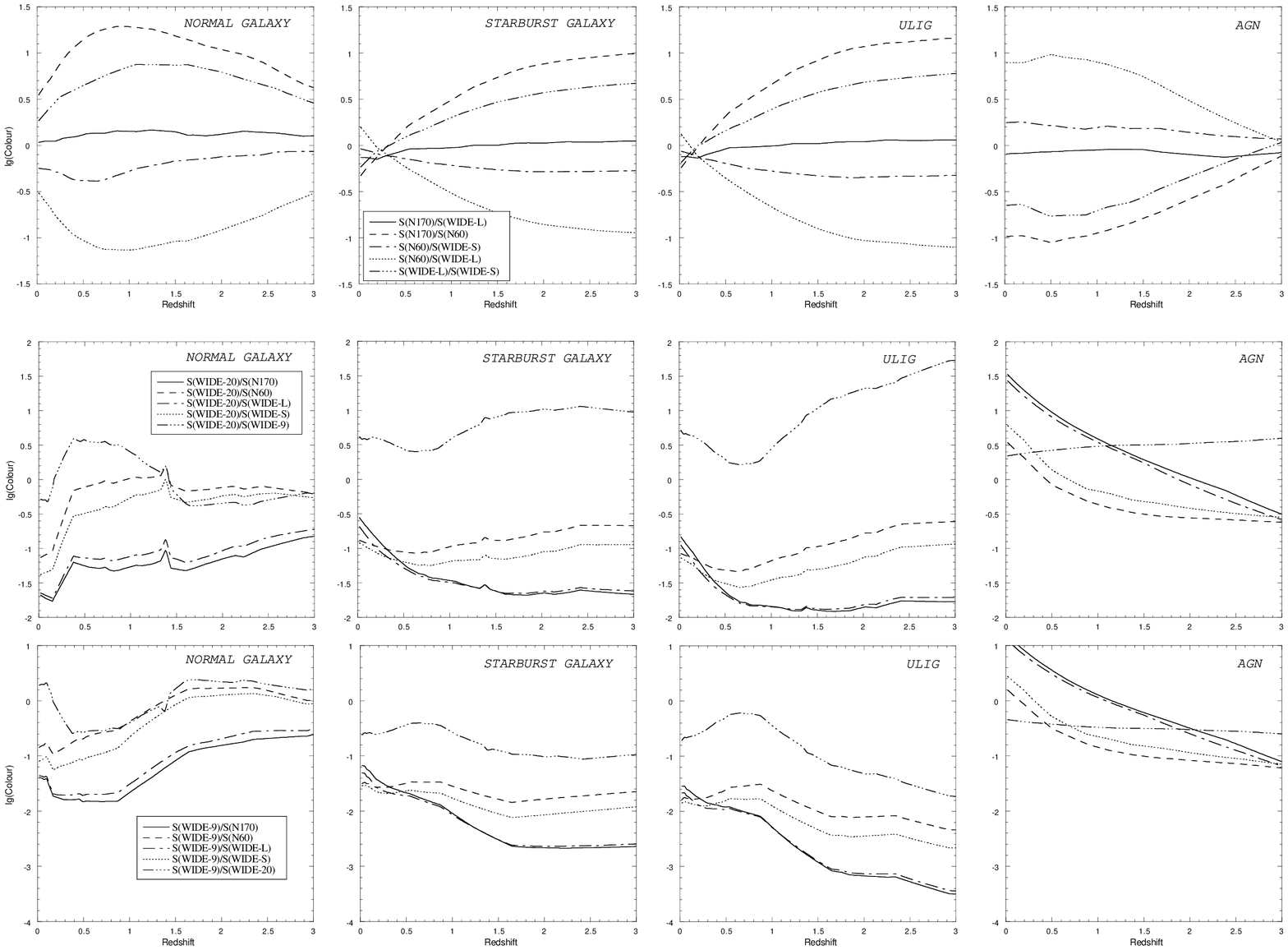,width=170mm}
}
\caption{The variation of galaxy colours with redshift calculated for the ASTRO-F All Sky Survey FIS bands ({\it top panels\,}) and MIR WIDE-20$\umu$m \& WIDE-9$\umu$m bands ({\it middle panels\,} \& {\it bottom panels\,} respectively). Colour combinations as given in the legends for a normal galaxy of $L_{60\mu m}=10^{10} L\sun $, a starburst galaxy of  $L_{60\mu m}=10^{11} L\sun $, an ultraluminous galaxy of $L_{60\mu m}= 10^{12} L\sun $ and an AGN of $L_{12\mu m}=10^{12} L\sun $.
\label{ccz}}
\end{figure*}

\subsection{The Visibility Function}
	The ASTRO-F All Sky Survey will be an invaluable tool in the study of large scale structure. The IRAS survey mapped the local large scale structure of the Universe out to redshifts $\approx$0.01-0.1 in a series of studies including 35000sq.deg.down to 600mJy at 60$\umu$m and 20000sq.deg. down to 200-250mJy at 12,25,60$\umu$m (e.g. Lonsdale et al.~\shortcite{lonsdale90}, Rush et al. ~\shortcite{rush93}, Oliver et al. ~\shortcite{oliver96}, Shupe et al. ~\shortcite{shupe98}, Saunders et al.~\shortcite {saun00}) using the IRAS PSC~\cite{beichman88} and FSS ~\cite{moshir91} data sets. Wide sky coverage is a necessity for large scale structure studies, as studies on smaller scales can be effected by localized fluctuations.  ASTRO-F will probe to 10-100 times deeper than IRAS over 500x the area of the SIRTF-SWIRE program ~\cite{lonsdale01}. A detailed analysis of the usefulness of ASTRO-F in the study of large scale structure is beyond the scope of this present work and will be addressed in a later paper; in this work, we briefly outline the potential of such studies using the All Sky Survey data. We approach our assessment via the density function and derived visibility function of galaxies in the ASTRO-F All Sky Survey (see Von Hoerner~\shortcite{vonhoerner73} \& Condon~\shortcite{condon84} for a more detailed derivation and explanation of the visibility function). In general, the total number of sources observable down to some limit $S$ is given by,
 \begin{equation}
  N(S)=\!\!\int_{0}^{\infty}\!\! \  \phi (L)  {\rmn{d}}lgL \!\!\int_{0}^{z}\!\! {\rmn{d}}V(z),
\end{equation}
where $\phi$ is the local differential luminosity function per decade in luminosity $L$. Integrating over the volume element ${\rmn{d}}V$ out to some limiting redshift $z$, the total number of sources. Discarding the second half of the integral gives $\!\!\int_{0}^{\infty}\!\! \  \phi (L)  {\rmn{d}}lgL$ gives the number density of galaxies in a given redshift bin at redshift $z$. This is effectively the galaxy luminosity function in a bin of width $dz$ centred on a redshift $z$, we call this the density function $\rho (z)$. The density function for the ASTRO-F All Sky Survey is shown in fig.~\ref{vis}a for all the FIS and MIR bands. The density function is essentially a measure of the survey sensitivity and plotting the corresponding results for the SIRTF-SWIRE survey the superiority of the latter's sensitivity can be clearly seen.

By taking logarithmic steps in the density, a threshold redshift $z_{o}(\rho _{o})$ can be read from the density function in fig.~\ref{vis}a. Using this redshift, $z_{o}$, a corresponding cosmological volume is calculated which will be the sampled volume above a given galaxy density for the proposed survey, where $V(\rho _{o})=V(z_{o})$. Convolving this volume with the total coverage of the survey gives us the quantity we refer to as the visibility function of sources in the all sky survey and is a function of both the depth (sensitivity) and the coverage of any given survey. The visibility function is plotted in fig.~\ref{vis}b and clearly shows the power and potential of the ASTRO-F All Sky Survey for large scale structure studies with 2-3 orders of magnitude in the volume being sampled at densities of $lg(\rho )=-7 \sim -4 Mpc^{-3}$ compared to the equivalent SIRTF-SWIRE bands. In truth, the ASTRO-F and SIRTF-SWIRE surveys supplement rather than supplant one another and are extremely complementary with ASTRO-F covering scales from $\sim$10-1000$h^{-1}$Mpc from z$\sim$0.01-1 and the SIRTF-SWIRE survey covering the range from scales of $\sim$10-100$h^{-1}$Mpc from redshifts $>$0.1-5.

\bigskip

\section{Galaxy Colours in the All Sky Survey }\label{sec:colours}

	Galaxy surveys at a single wavelength provide insight into source density and variation, background contributions and serendipitous discoveries but are by definition rather monochromatic. ASTRO-F will survey in 4 FIR (+2 MIR) bands allowing multi-band correlations and FIR (+MIR) colours producing as many as 6 new points on the IR SEDs of galaxies and source counts as a function of galaxy type (via colour discrimination similar to the way the IRAS galaxies were segregated into cirrus, starburst and AGN populations on the basis of their IR colours~\cite{mrr89}). Moreover, it may be also be possible to obtain rough photometric redshifts from the FIR data alone.

In fig.~\ref{ccz} the variation of galaxy colours with redshift calculated for the ASTRO-F All Sky Survey FIS bands ({\it top panels\,}) and MIR WIDE-20$\umu$m \& WIDE-9$\umu$m bands ({\it middle panels\,} \& {\it bottom panels\,} respectively), for a normal galaxy of $L_{60\mu m}=10^{10} L\sun $, a starburst galaxy of  $L_{60\mu m}=10^{11} L\sun $, an ultraluminous galaxy of $L_{60\mu m}= 10^{12} L\sun $ and an AGN of $L_{12\mu m}=10^{12} L\sun $. 

\begin{figure}
\centering
\centerline{
\psfig{figure=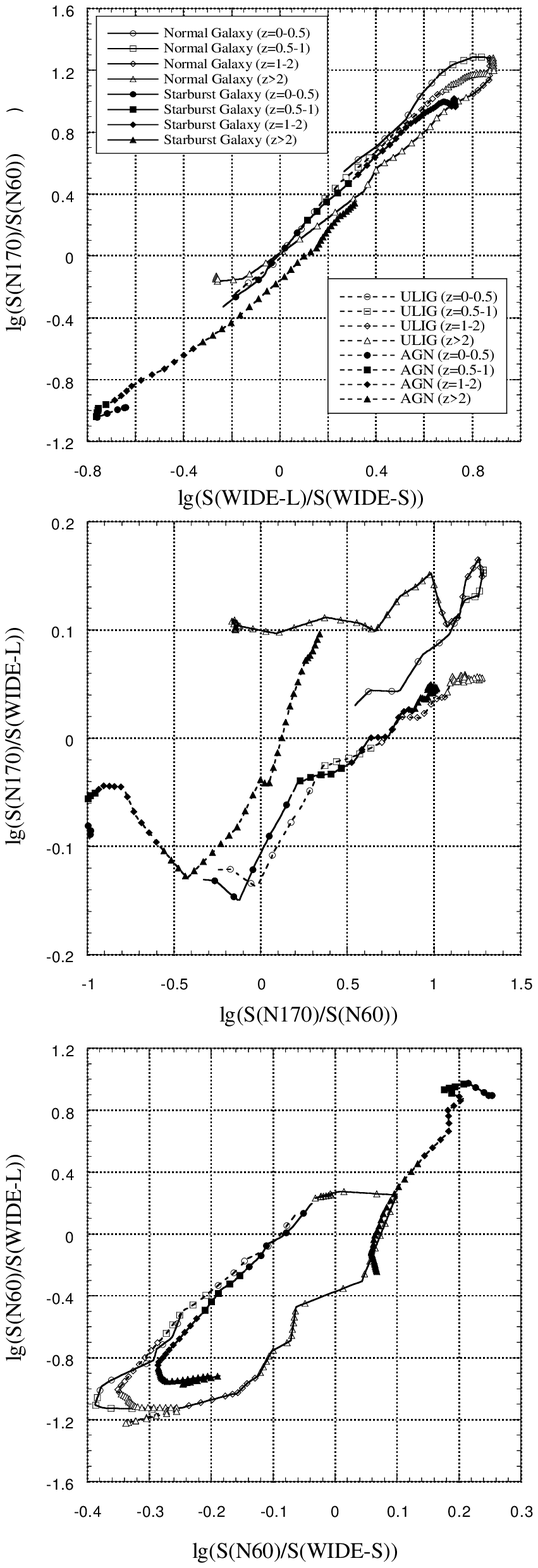,width=9cm}
}
\caption{Colour-Colour Diagrams for ASTRO-F All Sky Survey in the FIS bands derived from results in Fig.~\ref{ccz}. Results for normal, starburst, ULIG and AGN components are shown. Markers correspond to redshift steps of $\delta$z=0.1.
\label{cc}}
\end{figure}

For the normal and starburst (including LIG \& ULIG sources) galaxies, there is a general increase of the long wavelength (N170 \& WIDE-L) flux as the SED climbs through the Rayleigh-Jeans region resulting in redder N170/N60 \& WIDE-L/WIDE-S colours and bluer N60/WIDE-L colours. As the long wavelength band moves over the dust hump in the SED, the colours decrease in magnitude. For a normal galaxy the dust emission peaks at $\sim$100$\umu$m while the emission peak for starburst galaxies is at the shorter 60$\umu$m wavelength due to the higher dust temperatures in these galaxies. Thus, this effect occurs at z$\sim$0.7 and $\sim$1.9 for normal and starburst galaxies respectively. Since the dust torus SED we use for our AGN template is rather flat from 3-30$\umu$m and declines steeply to longer wavelengths, the short wavelength band flux must always be higher than the corresponding longer wavelength fluxes at least out to z$\approx$1, resulting in an inversion of the AGN colour variation with redshift compared to the other galaxy populations. Although this may seem at first sight a useful discriminator, one implication may be that the highest redshift AGN will in fact have similar colours to the low redshift normal galaxy population (see fig.~\ref{cc}). Previous work with IRAS/ISO data has also shown that Seyfert galaxies are indistinguishable from normal galaxies in 100-60$\umu$m colours or 200-100$\umu$m colours~\cite{spinoglio02}. However the ASTRO-F N170 band may provide additional discriminators that will allow us to unravel the 2 populations somewhat.

\begin{figure}
\centering
\centerline{
\psfig{figure=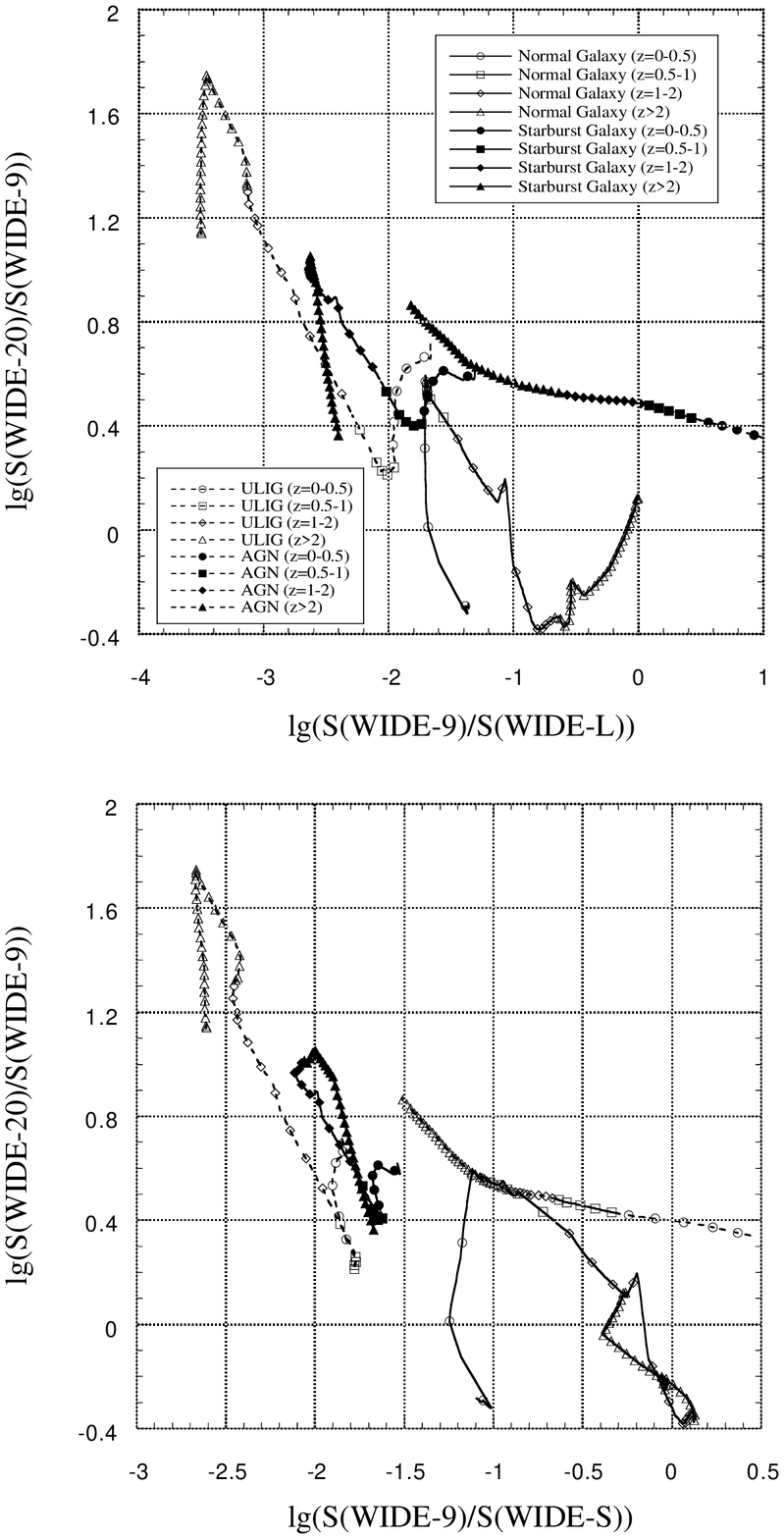,width=9cm}
}
\caption{Colour-Colour Diagrams for ASTRO-F All Sky Survey in the MIR+FIS bands derived from results in Fig.~\ref{ccz}. Results for normal, starburst, ULIG and AGN components are shown. Markers correspond to redshift steps of $\delta$z=0.1.
\label{ccmir}}
\end{figure}

The addition of the 2 wide MIR bands to the ASTRO-F All Sky Survey may provide further constraints on the colours of galaxies (bottom 2 panel rows of fig.~\ref{ccz}). However, their value as a statistical tool (for example to collect a large sample of photometric redshifts) will be severely limited by their low sensitivity, restricting such studies to z$<$0.1 where there would be $10^{4}$ sources over the entire sky corresponding to 1/700 and 1/90 N170 \& N60 sources respectively. In general, the low redshift starburst population will have decidedly redder 20$\umu$m/$\lambda$ colours than the normal galaxies. However, the colours of the low redshift AGN (z$<1$) will be significantly redder (especially in 20$\umu$m/N170, 20$\umu$m/WIDE-L, 9$\umu$m/N170, 9$\umu$m/WIDE-L) than all other sources. Again the is due essentially to the flat nature of the AGN SED from 3-30$\umu$m and may be useful as a discriminator of low redshift AGN.

\begin{table*}
\caption{Photometric survey follow up of ASTRO-F sources: Surveys (excluding IRAS) covering $\geq$1sr.}
\renewcommand{\arraystretch}{1.4}
\setlength\tabcolsep{15pt}
\begin{tabular}{@{}lllll}
\hline\noalign{\smallskip}
Survey & Reference & Area & Band & Sensitivity \\
\noalign{\smallskip}
\hline
\noalign{\smallskip}
POSS-II & Reid et al. 1991  & Northern Sky & optical &  B$_{j}$=22.5, R$_{c}$=20.8, I$_{c}$=19.5 \\
UKSS & Yentis et al. 1992 & Southern Sky & optical & b$_{j}$=21 \\
2MASS & Jarret et al. 2000 & All Sky & 2$\umu$m & J=15 (1.6mJy), K$_{s}$=13.5 (2.9mJy) \\
ROSAT & Voges et al. 1999 & All Sky & X-ray & 9x10$^{-12}$erg cm$^{-2}$ s$^{-1}$ \\
NVSS & Condon et al. 1998  & 82$\%$ Sky & radio & 2.5mJy \\
GALEX & Bianchi et al. 1999 & All Sky & UV & 20mJy (M$_{AB}$=20-21)\\
SDSS & Stoughton et al. 2002  & 10000sq.deg. & optical & u=22,g=22.2,r=22.2,i=21.3,z=20.5 \\
VISTA & www.vista.ac.uk & 10000sq.deg.& optical & U=25.6,B=26.7,V=26.5,R=25.3,I=24.7\\
VISTA & www.vista.ac.uk & 5000sq.deg.& NIR & J=23.4,H=21.9,K=20.5\\
UKIDSS & Warren \& Hewitt 2002 & 4000sq.deg.& NIR & Y=20.5, J=20, H=18.8, K=18.4\\
\noalign{\smallskip}
\hline
\noalign{\smallskip}
\end{tabular}\\
\label{followup}
\end{table*}

\begin{table*}
\caption{Redshift survey follow up of ASTRO-F sources}
\renewcommand{\arraystretch}{1.4}
\setlength\tabcolsep{15pt}
\begin{tabular}{@{}llll}
\hline\noalign{\smallskip}
Survey & Reference & Area & Band \& Sensitivity  \\
\noalign{\smallskip}
\hline
\noalign{\smallskip}
PSCz & Saunders et al. 2000  & All Sky & 60$\umu m>0.6$Jy\\
SDSS & Stoughton et al. 2002  & 10000sq.deg. & r$<$18.1\\
2dFGRS & Colless et al. 2001  & 2000sq.deg. & b$_{j}<19.45$\\
6dFGRS & Colless 2002  & 17000sq.deg. & K$<13$, H$<13.5$, J$<14.1$, I$<15.0$, B$<16.5$, S$_{60}>0.3Jy$\\
AA$\Omega$ 2df & www.ast.cam.ac.uk/AAO & 2deg.FOV & B$<$23 \\
\noalign{\smallskip}
\hline
\noalign{\smallskip}
\end{tabular}\\
\label{redsurvey}
\end{table*}

\begin{table*}
\caption{Median fluxes of ASTRO-F sources in the N60 band and from radio - X-ray wavelengths. These median fluxes are used to calculate the corresponding number of ASTRO-F beams per source to highlight the question of identification of ASTRO-F sources with follow up at K, I, B bands, radio wavelengths at 1.4GHz and X-ray ROSAT (0.5-2)keV band. Inverting the beams per source gives the number of sources per ASTRO-F beam where a number greater than unity will imply problematical identifications}
\renewcommand{\arraystretch}{1.4}
\setlength\tabcolsep{8pt}
\begin{tabular}{@{}llllllllllll}
\hline\noalign{\smallskip}
  & \multicolumn{6}{c}{Median Flux} &  \multicolumn{5}{c}{ASTRO-F beams per source} \\
Component  & N60 & K & I & B & Radio & X-ray &  K & I & B & Radio & X-ray \\
  & mJy & & & & mJy& erg/cm$^{2}$/s&  & & & & \\
\noalign{\smallskip}
\hline
\noalign{\smallskip}
Normal    & 73 & 12.2 & 14.0 & 15.8 & 0.5 & 1.1x10$^{-14}$ & 8330 & 8330 & 16600 & 180 & 360\\
Starburst & 67 & 16.3 & 16.2 & 19.5 & 0.7 & 0.6x10$^{-14}$ & 42 & 512 & 260 & 230 & 170\\
ULIG      & 58 & 17.9 & 19.9 & 22.0 & 0.6 & 0.5x10$^{-14}$ & 6 & 13 & 24 & 230 & 160\\
AGN       & 61 & 13.3 & 19.0 & 15.0 & 0.01 & 90x10$^{-14}$ & 1750 & 2220 & 11100 & 4 & 166660\\
\noalign{\smallskip}
\hline
\noalign{\smallskip}
\end{tabular}\\
\label{identifcation}
\end{table*}

Initial redshift estimates for the ASTRO-F All Sky Survey will be made via associations with photometric surveys at other wavelengths and extensive use of existing optical/near-IR redshift surveys and catalogues. In table~\ref{followup} \& table~\ref{redsurvey} we list large scale photometric surveys at other wavelengths and 4 of the largest present or near future large area redshift surveys respectively. These are of particular interest since it is in the domain of large area / volume statistics that ASTRO-F/SIRTF missions will be most discriminated from each other. Hirashita et al.~\shortcite{hirashita99} have suggested that typical ASTRO-F galaxies will have magnitudes B$<$21.2 \& H$<$19.6 meaning that they will be detectable in the POSS-II and UKSS catalogues (B$<$22.5, R$<$21.5, I$<$19) although they may be too faint for the SDSS (r$<$18.1) and 2dfGRS (B$_{j} <$19.45) redshift surveys. The PSCz \& SDSS comprise of 15,411 \& 10$^{6}$ galaxy redshifts respectively while the 2df has presently obtained more than 200,000 ~\cite{madgewick02}. However, these surveys are limited to z$<$0.02, 0.4, 0.2 respectively and additional redshift surveys / follow up may be utilized (e.g. 6df ~\cite{colless02}, AA$\Omega$2df, VLT-VIRMOS ~\cite{lefevre00}).

	The ASTRO-F All Sky Survey will produce a catalogue of sources numbering in the millions or even tens of millions. Identification of these sources may prove to be elusive in some cases. In this work we make a very brief initial analysis of the identification problem that ASTRO-F may face. Similar problems are encountered by many FIR-Submm facilities (a prime example being the difficulty in identifying SCUBA sources at other wavelengths due to the size of the SCUBA beam $~\sim 15''$ \cite{lilly99}, ~\cite{smail00}) since it is very difficult to identify sources at these wavelengths when the source densities are greater than 1 per beam (or conversly one beam/source). In table ~\ref{identifcation} we tabulate the estimated median fluxes of ASTRO-F galaxies in the N60 band. Using a sample of  template SEDs and the flux ratios between 60$\umu$m and other wavebands from radio to X-rays (Helou et al.~\shortcite{helou85}, Condon et al.~\shortcite{condon91}, White \& Becker~\shortcite{white92}, Rowan-Robinson et al.~\shortcite{mrr87}, Corbelli et al.~\shortcite{corbelli91}, Pearson \& Rowan-Robinson~\shortcite{cpp96},  Soifer et al. ~\shortcite{soifer86}, Green et al.~\shortcite{green92}) we calculate the median fluxes of these sources at 1.4GHz, Johnson K, I, B bands and the ROSAT 0.5-2keV X-ray band. 
Immediately by comparison with table~\ref{followup} we see that normal galaxies will be easily identified by both 2MASS or POSS-II/UKSS with around 1 misidentification in $\approx$2000 sources. These normal galaxies are quite bright and are predicted to be relatively local with their median redshift being $\approx$0.06. The starburst galaxies are also readily identifiable with POSS-II/UKSS, although we predict a slightly brighter B-band flux of 19.5 compared to the B$<$21.2 of Hirashita et al. ~\shortcite{hirashita99}. However, the LIG/ULIG sources will be more problematical being too faint for 2MASS, NVSS and possibly POSS-II/UKSS as well. The majority of the AGN detected by ASTRO-F should be picked up by the Hamburg-ESO Quasar survey in the area it covers ~\cite{wisotski00}.  We then predict and tabulate in table ~\ref{identifcation} the number of ASTRO-F beams per source at these wavelengths and flux levels using published surface densities of sources from the literature (compilations from Condon~\shortcite{condon84}, Afonso et al.~\shortcite{afonso01}, Fioc \& Rocca-Volmerange~\shortcite{fioc99}, Boyle et al.~\shortcite{boyle94}, McHardy et al.~\shortcite{mchardy98}). From table~\ref{identifcation} we see that if they are bright enough to be detected, most sources should be readily identifiable, i.e., more than a few beams per source. However the ULIGs will be elusive in the K,I,B bands as will the AGN in the radio. We would stress that this is a rather preliminary analysis of the identification problems and procedures that will be encountered with the ASTRO-F All Sky Survey sources. More sophisticated analysis, such as likelihood ratio IDs ~\cite{rutledge00} are beyond the present scope of this work and will be considered at a later time.

In addition, photometric redshifts from the 4 FIS-FIR bands (for sources z$>$0.1) plus an additional 2 MIR bands for sources z$<$0.1 will be used to provide rough constraints on the redshifts of sources. In figures~\ref{cc} \& ~\ref{ccmir} we plot the colour-colour diagrams for FIS bands and FIS+IRC bands respectively. We find in general that different galaxy types occupy distinct regions of the colour-colour plane and that by combining one or more of these plots that we should be able to obtain some degree of segregation in the galaxy populations both in type and in redshift. To illuminate some examples, for starburst, LIG \& ULIG sources there is a trend towards redder  N170/WIDE-L - N170/N60 colours with redshift with galaxies at z$>$2 having colours an order of magnitude redder than those at z$<$1. The N60/WIDE-L - N60/WIDE-S  colour-colour distribution of normal galaxies are similarly elongated with increasing redshift. However, as stated previously, there is some degeneracy between the low redshift colours of normal galaxies and the high redshift colours of starburst/LIG/ULIG sources. None of the colour-colour plots in fig.~\ref{cc} are capable of confidently separating these source populations from each other. The AGN, however, are very well segregated in both redshift and from the other galaxy populations by their colours. The low redshift AGN occupy very distinct areas in both the N60/WIDE-L- N60/WIDE-S and N170/N60 -  WIDE-L/WIDE-S colour-colour diagrams. It is a possibility that the colours of high redshift AGN may be confused with with the colours of high redshift normal (cool) galaxies although it is dubious that any any such cool sources could be detected in the ASTRO-F All Sky Survey since a normal galaxy at redshift $\sim$1 would have to have an IR luminosity $>10^{12}L_{\sun}$ to be detected in the all sky survey (see fig.~\ref{lz}). Therefore it seems reasonable to assume that any sources with lg(S(N60)/S(WIDE-L))$>\sim$0 in the N60/WIDE-L - N60/WIDE-S plane would indeed be high redshift AGN.

Combining the FIS-FIR colours in fig.~\ref{cc} with the IRC-MIR colours we find that we can obtain a set of more powerful discriminating tools. In fig. ~\ref{ccmir} we additionally plot the  WIDE-20/WIDE-9 - WIDE-9/WIDE-L \& WIDE-20/WIDE-9 - WIDE-9/WIDE-S colour-colour diagrams (Note that swapping the 2 MIR bands in these colour-colour plots will simply reflect the shape of the plots in the x-axis plane). With the addition of the MIR bands we find we can break the degeneracy between the low-z cool objects and high-z star bursting objects. In these diagrams, the normal galaxies lie blueward of the starburst populations in WIDE-20/WIDE-9 colours and in general have redder  WIDE-9/FIR colours than the star forming populations. Although the WIDE-9/FIR colours in both the low-z normal galaxies and high-z star bursting objects are both approximately constant and unchanging (since both the FIR flux the MIR-NIR flux is on a black body curve for both low-z normal and high-z starburst galaxy populations), the 20$\umu$m flux is moving through the UIB emission features while the 9$\umu$m flux climbs the normal galaxy SED while falling down through the starburst SED at higher redshifts causing a degeneracy between the 2 galaxy populations.

\bigskip

\section{Conclusions}\label{sec:conclusions}

	We have reviewed the ASTRO-F All Sky Survey and its expected results. ASTRO-F will carry out the first all sky survey at infrared wavelengths for 20 years and will probe down to sensitivities between 10-1000 times deeper than the IRAS mission to more than 5 times the angular resolution in 4 far-IR bands from 50-200$\umu$m and 2 mid-IR bands at 9 \& 20$\umu$m. Expected numbers of sources range from 10's millions in the longest wavelength bands where they will be source confusion limited, to 10,000 at mid-IR wavelengths (a number of an order of magnitude higher than that detected by IRAS or ISO at mid-IR wavelengths). Most of the sources detected in the far-infrared bands will be dusty LIGs \& ULIGs of which, in the longest wavelengths bands, $>$50$\%$ will be at redshifts greater than unity. 

The ultimate goal of the ASTRO-F All Sky Survey is to achieve approximately 95$\%$ sky coverage by the end of the survey to a reliability of $>$99$\%$ for bright sources. Absolute flux uncertainties of 10 and 20$\%$ are expected for point and diffuse sources respectively. Pointing accuracy is hoped to be $<5''$. Note that the orbit of ASTRO-F will be sun-synchronous polar with the  FOV always perpendicular to the Sun. This will have significant effects on the visibility of sources. Sources near the ecliptic plane will be visible only 2 days (29 orbits) every half year whereas targets near the ecliptic poles will be will be observable on many orbits. The North Ecliptic Pole is indeed under investigation as a target area for deep pointed surveys with the IRC \cite{cpp01a} and will complement well with the proposed Planck Deep NEP survey. Ultimately, it is expected that the ASTRO-F All Sky Survey will produce several catalogue products which could then serve as input catalogues and cross correlation databases for later missions such as Herschel \cite{pilbrat00} \& Planck (including the all sky survey with HFI ~\cite{coburn99}) due for launch in 2007. these catalogues can be broken down as follows;

\begin{enumerate}
  \item {\it ASTRO-F Flux of known sources} - Measurement of sources used for the Input Source Catalogue and fluxes of IRAS point sources.
  \item {\it Bright Source Catalogue} - A catalogue of sources that can be extracted easily such as bright sources or those at high ecliptic latitude.
  \item {\it Faint Source Catalogue} -  The all sky point source catalogue to the optimum sensitivity.
  \item {\it Image maps} - Crude maps, square degree imagelets and 10's sq.deg. image atlases.
\end{enumerate}

The ASTRO-F surveyor mission should be seen as supplementing not supplanting the SIRTF observatory mission. The two are complementary. SIRTF will cover relatively small areas (1-70sq.deg.) to high sensitivity while ASTRO-F will cover huge areas to more moderate sensitivities. In the study of large scale structure SIRTF-SWIRE will be sensitive to small scales over large redshift while conversely ASTRO-F will probe larger scales to shallower depths. However, one of many clear distinctions between the two missions is that the ASTRO-F All Sky Survey is the only way that the infrared dipole work carried out with IRAS can be checked and extended (Yahil et al.~\shortcite{yahil86}, Rowan-Robinson et al.~\shortcite{mrr00}, Rowan-Robinson et al.~\shortcite{mrr90b}).

ASTRO-F will also provide an ideal companion to PLANCK which will survey the whole sky at sub-mm wavelengths and also down to 200$\umu$m which will provide an important longer wavelength channel to join FIR measurements by ASTRO-F. Further associations and cross correlations with large scale surveys at other wavelengths (SWIRE, POS-II, UKSS, SDSS, NVSS, GALEX, ROSAT, UKIDSS, 6df, 2MASS, LMT, etc., see tables ~\ref{followup}, ~\ref{redsurvey}) will result in a fully comprehensive unified view of galaxy evolution and star formation from radio-mm-IR-optical-UV-X-ray wavelengths.

\bigskip

\section{Internet access to simulated data}

	The simulations for the ASTRO-F All Sky Survey discussed in this paper can be accessed through the world wide web at $http://astro.ic.ac.uk/\sim cpp/astrof/$. Other information on the ASTRO-F mission can be found at $http://www.ir.isas.ac.jp/$.
\bigskip

\section{Acknowledgements}

	During this work, cpp was supported by a fellowship of The Royal Society. cpp would like to thank Steve Serjeant for reading through this manuscript and providing many useful comments.
\bigskip



\bsp 

\label{lastpage}


\begin{thebibliography}{7}
%
\addcontentsline{toc}{section}{References}

\bibitem[\protect\citename{Altieri et al. }1999]{altieri99} Altieri B. et al., 1999, AA, 343, L65

\bibitem[\protect\citename{Afonso et al. }2001]{afonso01} Afonso J., Mobasher B., Hopkins A., Cram L., 2001, ApSS, 276, 941

\bibitem[\protect\citename{Aussel et al. }1999]{aussel99} Aussel H., Cesarsky C.J., Elbaz D., Starck J.L., 1999, AA, 342, 313

\bibitem[\protect\citename{Balbi et al. }2000]{balbi00} Balbi A. et al., 2000, ApJ, 545, L1

\bibitem[\protect\citename{Barger, Cowie \& Sanders }1999]{barg99} Barger A.J.,Cowie L.L., Sanders D.B., 1999, ApJ, 518, L5

\bibitem[\protect\citename{Barger et al. }1998]{barg98} Barger A.J. et al., 1998, Nature, 394, 248

\bibitem[\protect\citename{Beichman et al. }1988]{beichman88} Beichman C.A., Neugebauer G., Haling H.J., Clegg P.E., Chester T.J., 1988, IRAS Catalogs and Atlases, Vol1: Explanatory Supplement (Pasadena:JPL)

\bibitem[\protect\citename{Biviano et al. }2000]{biviano00} Biviano et al., 2000, in: Mezure A., Le Fevre O., Le Brun V., eds., ASP Conf.Ser., Clustering at High Redshifts, Astron.Soc.Pac., San Fransisco, p.101

\bibitem[\protect\citename{Boulade et al. }1996]{boul96} Boulade O., 1996, AA, 315, L85

\bibitem[\protect\citename{Boyle et al. }1994]{boyle94} Boyle B., Shanks T., Georgantopoulos I., Stewart G.C., Griffiths R.E., 1994, MNRAS, 271, 648

\bibitem[\protect\citename{Benn et al. }1993]{benn93} Benn C.R., Rowan-Robinson M., McMahon R.G., Broadhurst T.J., Lawrence A., 1993, MNRAS, 263, 98

\bibitem[\protect\citename{Bertin et al. }1997]{bertin97} Bertin E., Dennefield M., Moshir M., 1997, AA, 323, 685

\bibitem[\protect\citename{Bianchi et al. }1999]{bianchi99} Bianchi L. et al., 1999, Mem.Soc.Astron.Ital., 70, 365 

\bibitem[\protect\citename{Blain et al. }1999]{blain99} Blain A.W.,Kneib J.-P., Ivison R.J., Smail I., 1999, ApJ, 302, 632

\bibitem[\protect\citename{Boyle et al. }1988]{boyl88} Boyle N., Shanks T., Peterson B.A., 1988, MNRAS, 235, 935

\bibitem[\protect\citename{Calzetti \& Heckman }1999]{calzetti99} Calzetti D., Heckman T.M., 1999, ApJ, 519, 27

\bibitem[\protect\citename{Chary \& Elbaz }2001]{chary01} Chary R., Elbaz D., 2001, ApJ, 556, 562

\bibitem[\protect\citename{Coburn \& Murphy }1999]{coburn99} Coburn D., Murphy J., 1999,IrAJ, 26, 109

\bibitem[\protect\citename{Cohen et al. }2000]{cohen00} Cohen J.G. et al., 2000, ApJ, 538, 29

\bibitem[\protect\citename{Colless et al. }2001]{colless01} Colless M. et al., 2001, MNRAS, 328, 1039

\bibitem[\protect\citename{Colless }2002]{colless02} Colless M., 2002, In: The 6df Galaxy Survey Workshop, Anglo-Austalian Observatory, 30-31 May, 2002

\bibitem[\protect\citename{Condon }1974]{condon74} Condon J.J., 1974, ApJ, 188, 279
\bibitem[\protect\citename{Condon }1984]{condon84} Condon J.J., 1984, ApJ, 284, 44

\bibitem[\protect\citename{Condon et al.}1991]{condon91} Condon J.J., Anderson M.L., Helou G., 1991, ApJ, 376, 95

\bibitem[\protect\citename{Condon et al. }1998]{condon98} Condon J.J., Cotton W.D., Greisen E.W., Yin Q.F., Perley R.A., Taylor G.B., Broderick J.J., 1998, ApJ, 115, 1693

\bibitem[\protect\citename{Corbelli, Salpeter \& Dicky }1991]{corbelli91} Corbelli E., Salpeter E.,  Dicky J.M., 1991, ApJ, 370, 49

\bibitem[\protect\citename{Dennefeld }2000]{dennefeld00} Dennefeld M., 2000, In: S. Cristiani, A. Renzini, R.E. Williams (eds.), Deep Fields, Proc.ESO/ECF/STScI Workshop, Garching, Germany, 2000,  Springer-Verlag, 2001, 43

\bibitem[\protect\citename{Dickenson }1998]{dickenson98} Dickenson M., 1998, In: M. Livio, S.M. Fall, P. Madau (eds.), The Hubble Deep Field, Baltimore, STScI, 219

\bibitem[\protect\citename{Dole et al. }2000]{dole00} Dole H. et al. 2000, In: Lemke D, Stickel M.K. (Ed.)ISO Surveys of a Dusty Universe, Springer--Verlag, 54

\bibitem[\protect\citename{Dole et al. }2001]{dole01} Dole H. et al., 2001, AA, 372, 364

\bibitem[\protect\citename{Dunne et al. }2000]{dunne00} Dunne L., eales S., Edmunds M., Ivison R., Alexander P., Clements D.L., 2000, MNRAS, 315, 115

\bibitem[\protect\citename{Efstathiou \&  Rowan-Robinson }2002]{esf02} Efstathiou A., Rowan-Robinson M., 2002, MNRAS, {\it submitted}

\bibitem[\protect\citename{Efstathiou \& Siebenmorgen }2000]{esf00} Efstathiou,A.,Siebenmorgen,R., 2000, MNRAS, 313, 734

\bibitem[\protect\citename{Efstathiou et al. }2000a]{esf00a} Efstathiou,A. et al., 2000, MNRAS 319, 1169

\bibitem[\protect\citename{Elbaz et al. }1999]{elbaz99} Elbaz D. et al., 1999, AA, 351, L37

\bibitem[\protect\citename{Elbaz et al. }2000]{elbaz00} Elbaz D, 2000, In: Lemke D, Stickel M.K. (Ed.)ISO Surveys of a Dusty Universe, Springer--Verlag, 121

\bibitem[\protect\citename{Elbaz et al. }2002]{elbaz02} Elbaz D., Cesarsky C.J., Chanial P. Aussel H., Franceschini A., Fadda D., Chary R.R., 2002, AA, 384, 848

\bibitem[\protect\citename{Fang et al. }1998]{fang98} Fang F., Shupe D.~L., Xu C., Hacking P.~B., 1998, ApJ, 500, 693

\bibitem[\protect\citename{Fazio et al. }1998]{fazio98} Fazio G.G. et al., 1998, In: Fowler A.M. (Ed.)Infrared Astronomical Instrumentation, Proc. SPIE 3354, 114

\bibitem[\protect\citename{Finkbeiner et al. }2000]{fink00} Finkbeiner D.~P., Davis M., Schlegel D.~J., 2000, ApJ, 544, 81

\bibitem[\protect\citename{Fioc \& Rocca-Volmerange }1999]{fioc99} Fioc M., Rocca-Volmerange B., 1999, AA, 351, 869

\bibitem[\protect\citename{Fixsen et al. }1998]{fixsen98} Fixsen D.J. Dwek E., Mather J.C., Bennet C.L., Shafer R.A., 1998, ApJ, 508, 123

\bibitem[\protect\citename{Flores et al. }1999a]{flores99a} Flores H.et al., 1999a, ApJ, 517, 148

\bibitem[\protect\citename{Flores et al. }1999b]{flores99b} Flores H. et al., 1999b, AA, 343, 389

\bibitem[\protect\citename{Fox et al. }2002]{fox02} Fox M. et al., 2002, MNRAS, 331, 839

\bibitem[\protect\citename{Franceschini et al. }2001]{franceschini91} Franceschini A., Toffolatti L., Mazzei P., Danese L., De Zotti G., 1991, AASS, 89, 285

\bibitem[\protect\citename{Freedman et al. }2001]{freedman01} Freedman W.L. et al., 2001, ApJ, 553, 47

\bibitem[\protect\citename{Gardner et al. }2000]{gardner00} Gardner J.P. et al., , 2000, AJ, 119, 486

\bibitem[\protect\citename{Goldader et al. }2002]{goldader02} Goldader J.D., Meurer G., Heckman T.M., Seibert M., Sanders D.B., Calzetti D., Steidel C.,  2002, ApJ, in press

\bibitem[\protect\citename{Gorijian, Wright \& Chary }2002]{Gorijian00} Gorijian V., Wright E.L., Chary R.R., 2000, ApJ, 536, 550

\bibitem[\protect\citename{Green et al. }1992]{green92} Green P.J., Anderson S.F., Ward M.J., 1992, MNRAS, 254, 30

\bibitem[\protect\citename{Gregorich et al. }1995]{gregor95}Gregorich D.T., Neugebauer G., Soifer B.T., Gunn J.E., Herter T.L., 1995, AJ, 110, 259

\bibitem[\protect\citename{Gruppioni et al. }2000]{gruppioni99} Gruppioni C. et al., 1999, MNRAS, 305, 297

\bibitem[\protect\citename{Hacking \& Houck }1987]{hacking87} Hacking P.B., Houck J.R., 1987, ApJS, 63, 311

\bibitem[\protect\citename{Hauser et al. }1998]{hauser98} Hauser M.G. et al., 1998, ApJ, 508, 25

\bibitem[\protect\citename{Hauser }2001]{hauser01} Hauser M.G., (2000), In:  M. Harwit, M. Hauser (Eds), XXIVth IAU General Assembly, S204 The Extragalactic Infrared Background and its Cosmological Implications, 2001,  101

\bibitem[\protect\citename{Helou et al. }1984]{helou84} Helou G., Soifer B.T., Rowan-Robinson M., 1984, BAAS, 16, 471

\bibitem[\protect\citename{Helou et al. }1985]{helou85} Helou G., Soifer B.T., Rowan-Robinson M., 1985, ApJ, 298, L7

\bibitem[\protect\citename{Hirashita et al. }1999]{hirashita99} Hirashita H., Takeuchi T., Ohta K., Shibai H., 1999, PASJ, 51, 81

\bibitem[\protect\citename{Helou \& Beichman }1990]{helou90} Helou G., Beichman C.A., 1990, In: From Ground Based to Space Borne sub-mm Astronomy, Proc. of 29th Liege International Astrophysical Col., ESA Publ., 117

\bibitem[\protect\citename{Hogg }2001]{hogg01} Hogg D.W., 2001, AJ, 121, 1207

\bibitem[\protect\citename{Holland et al. }1999]{holland99} Holland W. et al., 1999, MNRAS, 303, 659

\bibitem[\protect\citename{Hughes et al. }1998]{hugh98} Hughes D. et al., 1998, Nature, 457, 616

\bibitem[\protect\citename{Ishihara et al. }2002]{ishihara02} Ishihara D. et al., 2002, Proc. SPIE

\bibitem[\protect\citename{Jarrett et al. }2000]{jarrett00} Jarrett T.H., Chester T., Cutri R., Schneider S., Skrutskie M., Huchra J.P., 2000, AJ, 119, 2498

\bibitem[\protect\citename{Kaneda et al. }2002]{kaneda02} Kaneda H., Onaka T., Yamashiro R., Nakagawa T., 2002, Proc. SPIE

\bibitem[\protect\citename{Kawada }1998]{kaw98} Kawada M., 1998 In: Fowler A.M. (Ed.)Infrared Astronomical Instrumentation, Proc. SPIE 3354, 905

\bibitem[\protect\citename{kawara et al. }1998]{kawara98} Kawara K. et al., 1998, AA, 336, L9

\bibitem[\protect\citename{Kessler et al. }1996]{kessler96} Kessler M., et al., 1996, AA, 315, L27

\bibitem[\protect\citename{Kim \& Sanders }1998]{kim98} Kim D.C., Sanders D.B., 1998, ApJS, 119, 41

\bibitem[\protect\citename{Kravtsov et al. }1998]{kravtsov98} Kravtsov A.V., Klypin A.A., Khokhlov A.M., 1998, BAAS, 30, 1306

\bibitem[\protect\citename{Lagache \& Puget }2000]{lagache00b} Lagache G., Puget J-L., 2000, AA, 355, 17L

\bibitem[\protect\citename{Lagache et al. }2000]{lagache00a} Lagache G., Haffner L.M., Reynolds R.J., Tufte S.L., 2000, AA, 354, 247

\bibitem[\protect\citename{Lawrence et al. }1986]{law86} Lawrence A., Walker D., Rowan-Robinson M., Leech K.J., Penston M.V., 1986, MNRAS, 219, 687

\bibitem[\protect\citename{Le Fevre et al. }2000]{lefevre00} Le Fevre O. et al., 2000, In: Cristiani S., Renzini R.E., Williams R.E., (Ed.) Deep Fields, Springer--Verlag, 2001, 236 

\bibitem[\protect\citename{Lilly et al. }1999]{lilly99} Lilly S.J. et al., 1999, ApJ, 518, 641 

\bibitem[\protect\citename{Lonsdale et al. }1990]{lonsdale90} Lonsdale C.J., Hacking P.B., Conrow T.P., Rowan-Robinson M., 1990, ApJ, 358, 60

\bibitem[\protect\citename{Lonsdale }2001]{lonsdale01} Lonsdale C.J., 2001, BAAS, 198, 2502

\bibitem[\protect\citename{Low et al. }1984]{low84} Low F.J. et al. 1984, ApJ, 278, L19

\bibitem[\protect\citename{Lu et al. }1996]{lu96} Lu N.Y.et al., 1996, BAAS, 28, 1356

\bibitem[\protect\citename{Madau et al. }1996]{madau96} Madau P., Ferguson H.C., Dickenson M., Giavalisco M., Steidel C., Fruchter A., 1996, MNRAS, 283, 1388

\bibitem[\protect\citename{Madgewick et al. }2002]{madgewick02} Madgewick D.S. et al., 2002, MNRAS, 333, 133

\bibitem[\protect\citename{Maihara et al. }2001]{maihara01} Maihara T. et al., , 2001, PASJ, 53, 25

\bibitem[\protect\citename{Mann et al. }2002]{mann02} Mann R.G. et al., 2002, MNRAS, astro-ph/0201510

\bibitem[\protect\citename{Matsuhara }1998]{mat98} Matsuhara H., 1998, In: Fowler,A.M. (Ed.)Infrared Astronomical Instrumentation, Proc. SPIE 3354, 915

\bibitem[\protect\citename{Matsuhara }2000]{mat00} Matsuhara H. et al., 2000, AA, 361, 407

\bibitem[\protect\citename{Matsumoto et al. }2000]{mats00} Matsumoto T. et al, 2000, In: Lemke D, Stickel M.K. (Ed.)ISO Surveys of a Dusty Universe, Springer--Verlag, 99

\bibitem[\protect\citename{McHardy et al. }1998]{mchardy98} McHardy I.M. et al.,  1998, MNRAS, 295, 641

\bibitem[\protect\citename{Meurer et al. }1999]{meurer99} Meurer G.R.. Heckman T.M., Calzetti D., 1999, ApJ, 521, 64

\bibitem[\protect\citename{Moshir et al. }1991]{moshir91} Moshir M., 1991, Explanatory Supplement to the IRAS Faint Source Survey, Version 2 (Pasadena:JPL)

\bibitem[\protect\citename{Murakami }1998]{mura98} Murakami H., 1998, In: Bely P.Y., Breckinridge J.B. (Ed), Space Telescopes and Instruments, Proc. SPIE 3356, 471

\bibitem[\protect\citename{Oliver et al. }1992]{oliver92} Oliver S.J., Rowan-Robinson M., Saunders W., 1992, MNRAS, 256, 15

\bibitem[\protect\citename{Oliver et al. }1996]{oliver96} Oliver S.J. et al., 1996, MNRAS, 280, 673

\bibitem[\protect\citename{Oliver et al. }1997]{oliver97} Oliver S.J. et al., 1997, MNRAS, 289, 471

\bibitem[\protect\citename{Oliver et al. }2002]{oliver02} Oliver S.J., 2002, MNRAS, astro-ph/0201506

\bibitem[\protect\citename{Pearson \& Rowan-Robinson }1996]{cpp96} Pearson C.P.,Rowan-Robinson M., 1996, MNRAS, 283, 174

\bibitem[\protect\citename{Pearson et al.}1997]{cpp97} Pearson C.P.,Rowan-Robinson M., McHardy I.M., Jones L.R., Mason K.O.,  1997, MNRAS, 288, 285

\bibitem[\protect\citename{Pearson et al. }2001]{cpp01a} Pearson C.P., Matsuhara H., Onaka T., Watarai H., Matsumoto T., 2001, MNRAS 324, 999

\bibitem[\protect\citename{Pearson }2001]{cpp01b} Pearson C.P. 2001, MNRAS 325, 1511 (cpp)

\bibitem[\protect\citename{Pettini et al. }1998]{pettini98} Pettini M., Kellog M., Steidel C., Dickenson M., Adelberger K.L., Giavalisco M., 1998, ApJ, 508, 539

\bibitem[\protect\citename{Pilbrat }2000]{pilbrat00} Pilbrat G.L., 2000, In: Lemke D, Stickel M.K. (Ed.)ISO Surveys of a Dusty Universe, Springer--Verlag, p.408

\bibitem[\protect\citename{Puget \& Leger }1989]{pug89} PugetJ.L., Leger A., 1989, ARAA, 27, 161

\bibitem[\protect\citename{Puget et al. }1996]{puget96} Puget J-L. et al., 1996, AA, 308, L5

\bibitem[\protect\citename{Puget et al }1999]{pug99} PugetJ.L. et al., 1999, AA, 345, 29

\bibitem[\protect\citename{Reid et al }1991]{reid91} Reid I.N. et al., 1991,PASP, 103, 661

\bibitem[\protect\citename{Rieke }2000]{rieke00} Rieke G.H., 2000, In: Lemke D, Stickel M.K. (Ed.)ISO Surveys of a Dusty Universe, Springer--Verlag, 403

\bibitem[\protect\citename{Rowan-Robinson et al. }1987]{mrr87} Rowan-Robinson M., Helou G., Walker D., 1987, MNRAS, 227, 589

\bibitem[\protect\citename{Rowan-Robinson \& Crawford }1989]{mrr89} Rowan-Robinson M., Crawford P. 1989, MNRAS, 238, 523

\bibitem[\protect\citename{Rowan-Robinson et al. }1990]{mrr90} Rowan-Robinson M., Hughes J., Vedi K., Walker D.W., 1990, MNRAS, 246, 273

\bibitem[\protect\citename{Rowan-Robinson et al. }1990b]{mrr90b} Rowan-Robinson M. et al., 1990, MNRAS, 247, 1

\bibitem[\protect\citename{Rowan-Robinson et al. }1991a]{mrr91a} Rowan-Robinson M., Saunders W., Lawrence A., Leech K., 1991a, MNRAS, 253, 485

\bibitem[\protect\citename{Rowan-Robinson et al. }1991b]{mrr91b} Rowan-Robinson M. et al., 1991, Nature, 351, 719

\bibitem[\protect\citename{Rowan-Robinson }1995]{mrr95} Rowan-Robinson M., 1995, MNRAS, 272, 737

\bibitem[\protect\citename{Rowan-Robinson et al. }1997]{mrr97} Rowan-Robinson M. et al., 1997, MNRAS, 289, 490

\bibitem[\protect\citename{Rowan-Robinson et al. }2000]{mrr00} Rowan-Robinson M. et al., 2000, MNRAS, 314, 375

\bibitem[\protect\citename{Rowan-Robinson }2000]{mrr00a} Rowan-Robinson M., 2000, MNRAS, 885, 900

\bibitem[\protect\citename{Rush, Malkan \& Spinoglio }1993]{rush93} Rush B., Malkan M., Spinoglio L., 1993, ApJSS, 89, 1

\bibitem[\protect\citename{Rutledge et al. }2000]{rutledge00} Rutledge R.E.,Brunner R.J., Prince T.A., Lonsdale C., 2000, ApJSS, 131, 335

\bibitem[\protect\citename{Sanders \& Mirabel }1996] {sand96} Sanders D.B., Mirabel I.F., 1996, ARAA, 34, 725

\bibitem[\protect\citename{Saunders }1990]{saun90p} Saunders W., 1990,  Ph.D. thesis, Queen Mary \& Westfield College, University of London

\bibitem[\protect\citename{Saunders et al. }1990]{saun90} Saunders W., Rowan-Robinson M., Lawrence A., Efstathiou G., Kaiser N., Ellis,R.S., 1990, MNRAS, 242, 318

\bibitem[\protect\citename{Saunders et al. }2000]{saun00} Saunders W.et al. 2000, MNRAS, 317, 55

\bibitem[\protect\citename{Scott et al. }2000]{scott00} Scott S.E. et al., 2000, MNRAS, 331, 817

\bibitem[\protect\citename{Serjeant et al. }2000]{serjeant00} Serjeant S.B.G. et al., 2000, MNRAS, 316, 768

\bibitem[\protect\citename{Shupe, Fang \& Hacking }1998]{shupe98} Shupe D.~L., Fang F., Hacking P.~B., 1998, ApJ, 501, 597

\bibitem[\protect\citename{Smail, Ivison \& Blain }1997]{smail97} Smail I., Ivison R.J., Blain A.W., 1997, ApJ, 490, L5

\bibitem[\protect\citename{Smail et al. }2000]{smail00} Smail I., Ivison R.J., Owen F.N., Blain A.W., Kneib J.P., 2000, ApJ, 528, 612

\bibitem[\protect\citename{Soifer et al. }1986]{soifer86} Soifer B.T. et al., 1986, ApJ, 303, L41

\bibitem[\protect\citename{Soifer et al. }1987]{soifer87} Soifer B.T., Houck J.R., Neugebauer G., 1987, ARAA, 25, 187

\bibitem[\protect\citename{Spinoglio et al. }2002]{spinoglio02} Spinoglio L., Andreani P., Malkan M.A., 2002, astro-ph/0202331

\bibitem[\protect\citename{Steidel et al. }1999]{steidel99} Steidel C., Adelberger K.L., Giavalisco M., Dickenson M., Pettini M., 1999, ApJ, 519, 1

\bibitem[\protect\citename{Steidel et al. }1996]{steidel96} Steidel C., Giavalisco M., Pettini M., Dickenson M., Adelberger K.L., 1996, ApJ, 462, L17

\bibitem[\protect\citename{Stoughton et al. }2002]{stoughton02} Stoughton C. et al., 2002, AJ, 123, 485

\bibitem[\protect\citename{Takagi et al. }2002]{takagi02} Takagi T., Arimoto N., Hanami N., 2002, MNRAS, submitted

\bibitem[\protect\citename{Takahashi et al. }2000]{takahashi00} Takahashi H. et al., 2000, In: Breckinridge J.B. (Ed), UV, Optical, and IR Space Telescopes and Instruments, Proc. SPIE 4013, 47

\bibitem[\protect\citename{Takeuchi et al. }1999]{take99} Takeuchi T.T., Hirashita H., Ohta K., Hattori G.T., Ishii T.T., Shibai H., 1999, PASP, 111, 288

\bibitem[\protect\citename{Totani \& Takeuchi }2002]{totani02} Totani T., Takeuchi T.T., 2002, ApJ, 570, 470

\bibitem[\protect\citename{Vigroux et al. }1996]{vig96} Vigroux L. et al., 1996, AA, 315, L93

\bibitem[\protect\citename{Voges et al. }1999]{voges99} Voges W. et al., 1999, AA, 349, 389

\bibitem[\protect\citename{Von Hoerner }1973]{vonhoerner73} Von Hoerner S., 1973, ApJ, 186, 741

\bibitem[\protect\citename{Wada et al. }2002]{wada02} Wada T. et al., 2002, Proc. SPIE

\bibitem[\protect\citename{Warren \& Hewitt }2002]{warren02} Warren S., 2002, astro-ph/0201216

\bibitem[\protect\citename{Watarai et al. }2000]{wat00} Watarai H. et al., 2000, In: Breckinridge J.B. (Ed), UV, Optical, and IR Space Telescopes and Instruments, Proc. SPIE 4013, p.59

\bibitem[\protect\citename{White \& Becker }1992]{white92} White R.L., Becker R.H., 1992, ApJSS, 79, 331

\bibitem[\protect\citename{Williams et al. }2000]{williams00} Williams R.E. et al., 2000, AJ, 120, 2735

\bibitem[\protect\citename{Williams et al. }1996]{williams96} Williams R.E. et al., 1996, AJ, 112, 1335

\bibitem[\protect\citename{Wisotski et al. }2000]{wisotski00} Wisotski L., Christlieb N., Bade N., Beckmann V., Kohler T., Vanelle C., Reimers D., 2000, AA, 358, 77

\bibitem[\protect\citename{Wright \& Rees }2000]{wright00} Wright E.L., Reese E.D., 2000, ApJ, 545, 43

\bibitem[\protect\citename{Xu et al. }1998]{xu98} Xu,C. et al. 1998, ApJ, 508, 576

\bibitem[\protect\citename{Xu et al. }2000]{xu00} Xu C., Lonsdale C.J., Shupe D.L., O'Linger J., Masci F. 2000, ApJ, 562, 179

\bibitem[\protect\citename{Yahil et al. }1986]{yahil86} Yahil A., Walker D., Rowan-Robinson M., 1986, ApJ, 301, L1

\bibitem[\protect\citename{Yentis et al. }1992]{yentis92} Yentis D.J., Cruddace R.G., Gursky H., Stuart B.V., Wallin J.F., MacGillivray H.T., Collins C.A., 1992, In:  MacGillivray H.T., Thomson E.B. (Ed), Digitised Optical Surveys, Kluwer, Dordrecht, 67 

\end{thebibliography}
\end{document}